\documentclass[amsmath,amssymb,prr,aps,showpieces,superscriptaddress,twocolumn,bibnotes,nobibnotes,floatfix]{revtex4-2}
\usepackage{amsmath,amsfonts,amssymb,amsthm,graphics,graphicx,epsfig,bbm}
\usepackage[colorlinks=true,citecolor=blue,linkcolor=blue,urlcolor=blue]{hyperref}
\usepackage[usenames]{color}
\usepackage{graphicx}
\usepackage{subfigure}
\usepackage{amsmath}
\usepackage{epsfig}
\usepackage{dcolumn}
\usepackage{bm}
\usepackage{color}
\usepackage{times}
\usepackage{epstopdf}
\usepackage[dvipsnames]{xcolor}
\usepackage{amssymb}
\usepackage{amstext}
\usepackage{latexsym}
\usepackage{comment}
\usepackage{hyperref}
\usepackage{amsfonts}
\usepackage{psfrag}
\usepackage{soul,xcolor}
\usepackage[normalem]{ulem}
\usepackage{dsfont}
\usepackage{txfonts}
\usepackage{float}
\usepackage{algorithm}
\usepackage{algpseudocode}
\usepackage{booktabs}
\usepackage{dblfloatfix}
\usepackage{capt-of}
\usepackage[percent]{overpic}
\usepackage[T1]{fontenc}

\usepackage{xcolor}

\newcommand{\ket}[1]{\vert #1 \rangle}
\newcommand{\bra}[1]{\langle #1 \vert}
\newcommand{\ketbra}[2]{\vert #1 \rangle \langle #2 \vert}

\newcommand{\Tr}{\mathrm{Tr}}

\usepackage{tikz,xcolor,hyperref}
\definecolor{lime}{HTML}{A6CE39}
\DeclareRobustCommand{\orcidicon}{%
	\begin{tikzpicture}
	\draw[lime, fill=lime] (0,0) 
	circle [radius=0.16] 
	node[white] {{\fontfamily{qag}\selectfont \tiny ID}};
	\draw[white, fill=white] (-0.0625,0.095) 
	circle [radius=0.007];
	\end{tikzpicture}
	\hspace{-2mm}
}

\usepackage{orcidlink} % en el preámbulo

\foreach \x in {A, ..., Z}{%
	\expandafter\xdef\csname orcid\x\endcsname{\noexpand\href{https://orcid.org/\csname orcidauthor\x\endcsname}{\noexpand\orcidicon}}
}

\renewcommand{\orcidB}[1]{\orcidlink{#1}}

\begin{document}

\setstcolor{red}

\title{Qudit-ADAPT-VQE: an adaptive variational algorithm with counterdiabatic-inspired improvements for qudits}
\date{\today}

\author{Joaquín Molina
\orcidB{0009-0002-7963-709X}}
\affiliation{Facultad de F\'isica, Pontificia Universidad Cat\'olica de Chile, Santiago 7820436, Chile}

\author{Herbert Díaz-Moraga
\orcidB{0009-0007-1446-9060}}
\affiliation{Facultad de F\'isica, Pontificia Universidad Cat\'olica de Chile, Santiago 7820436, Chile}

\author{Dardo Goyeneche
\orcidB{0000-0002-9865-4226}}
\affiliation{Facultad de F\'isica, Pontificia Universidad Cat\'olica de Chile, Santiago 7820436, Chile}

\author{Diego Tancara 
\orcidB{0000-0002-5053-3521}} 
\email{datancara@uc.cl}
\affiliation{Facultad de F\'isica, Pontificia Universidad Cat\'olica de Chile, Santiago 7820436, Chile}

\begin{abstract}    
Variational quantum algorithms based on qudits have attracted significant attention in recent years due to their potential for more efficient quantum information encoding and processing. However, as in their qubit-based counterparts, challenges such as barren plateaus and the design of efficient ansatzes remain major obstacles. In this work, we propose to address these issues through a qudit implementation of the ADAPT-VQE algorithm, which constructs the ansatz iteratively. Specifically, we introduce an operator pool inspired by adiabatic evolution enhanced with counterdiabatic driving for ansatz construction and employ it to solve the Max 3-Cut and Multiway Number Partitioning problems. We show that the warm-start strategy inherent to ADAPT-VQE, together with an ansatz construction based on counterdiabatic operators, achieves higher accuracy than fixed-ansatz approaches and, at equal number of variational parameters, requires fewer native gates. Furthermore, we show that, in qudit-based quantum computing, ADAPT-VQE with a counterdiabatic operator pool can navigate rough optimization landscapes with local traps through the burrowing mechanism, suggesting robustness against local minima and providing a flexible framework for variational quantum algorithms with qudits.

\end{abstract}

\maketitle

\section{Introduction}
In recent years, alternatives to qubit-based quantum computing have attracted increasing attention. The standard qubit paradigm has become the dominant framework for quantum computing, partly because it provides a natural quantum analogue of the binary logic underlying classical information processing. However, restricting quantum information to two-level units may not always provide the most resource-efficient representation, nor fully exploit the structure of available quantum hardware. In this context, advances in the controlled experimental implementation of multilevel quantum systems with more than two accessible states, known as qudits \cite{FisherPRXQuantum2023, HrmoNatCommun2023, KruckenhauserQST2023, LowNPJQuantumInf2025, WangPhysRevAppl2025, GossNPJQuantumInf2024, ChiNatCommun2022}, have motivated the development of quantum computing paradigms based on these systems rather than qubits. This progress has also been accompanied by experimental realization of qudit gate operations across different platforms, including trapped ions \cite{HrmoNatCommun2023, AksenovPRA2023}, superconducting circuits \cite{LuoPRL2023, GossNatCommun2022}, and high-dimensional photonic systems \cite{LiuNatPhoton2026, DuOptLaserTechnol2025}. Qudit-based quantum computing offers several potential advantages over the conventional qubit-based counterpart. Since each qudit is described by a $d$-dimensional local Hilbert space, with $d>2$, a given computational space can be represented using fewer quantum units than in a two-level encoding. This reduction may lower circuit complexity and improve resource efficiency~\cite{WangFrontPhys2020,GaoQuantum2023}, which is particularly relevant in the Noisy Intermediate Scale Quantum (NISQ) regime, where increasing the number of quantum units generally makes quantum processors more susceptible to noise, calibration overhead, and control limitations~\cite{ErhardNatCommun2019,DaiPRXQuantum2021}. Moreover, qudits provide a natural representation for problems formulated in variables with dimensionality higher than binary, such as Max $k$-Cut, Graph Coloring, or Multiway Number Partitioning, where each variable can naturally take more than two possible values. In this context, several optimization problems have been addressed using qudit-based variational quantum algorithms, with different implementations proposed for algorithms such as Quantum Approximate Optimization Algorithm (QAOA) for qudits \cite{DellerPRA2023, EkstromPRR2025} which is inspired by adiabatic quantum optimization and implements its evolution through a variational circuit, and counterdiabatic-inspired ansatz, where approximate counterdiabatic terms are used to enhance the variational search by suppressing diabatic excitations~\cite{BottarelliPRR2025}. Alternative qudit-based optimization strategies have also been proposed, including adiabatic quantum optimization~\cite{AminQIP2013,AngkhanawinQST2026} and counterdiabatic quantum optimization~\cite{TancaraNPJQuantumInf2025}. Nevertheless, variational quantum algorithms have become especially prominent due to their flexibility and compatibility with NISQ devices. Their hybrid quantum--classical structure allows one to tailor the ansatz, cost function, and optimization strategy to the problem and hardware under consideration.

Despite these advantages, variational quantum algorithms also face important limitations. In qubit-based quantum computing, it is now well established that the trainability of parametrized quantum circuits can be severely hindered by barren plateaus, where the cost-function gradients vanish exponentially with system size, making classical optimization inefficient \cite{LaroccaNatRevPhys2025}. Recent work has further shown that, for certain classes of qubit-based variational circuits, provable absence of barren plateaus can be accompanied by efficient classical simulability~\cite{Cerezo2025}. In addition, the performance of variational algorithms depends crucially on the choice of ansatz: shallow circuits may lack sufficient problem structure, whereas highly expressive ansatzes can become difficult to optimize and may require large circuit depths \cite{HolmesPRXQuantum2022}. This ansatz-design problem becomes particularly relevant in qudit-based quantum computing, where the enlarged local Hilbert space can require more expressive parametrizations while also amplifying barren plateaus problem, as recently shown in Ref.~\cite{FriedrichQMI2025}. Among the different strategies proposed to obtain efficient ansatzes with shallow circuits and sufficient expressivity in qubit-based quantum computing, the adaptive derivative-assembled problem-tailored variational quantum eigensolver (ADAPT-VQE) \cite{GrimsleyNatCommun2019} has attracted growing attention in recent years. Unlike fixed ansatz approaches, ADAPT-VQE constructs the ansatz iteratively by selecting, at each step, the operator that provides the largest energy gradient from a predefined operator pool. This procedure yields a problem-tailored ansatz that incorporates only the most relevant directions in Hilbert space, reducing circuit depth and avoiding unnecessary variational parameters. In addition, its gradient-guided construction has been shown to provide robustness against rough parameter landscapes and barren plateau effects: by adding one operator at a time and reusing the optimized parameters from the previous iteration in a warm-start way, ADAPT-VQE remains localized in regions with resolvable gradients and can progressively “burrow” through local traps toward the solution \cite{GrimsleynpjQuantumInf2023}. Although ADAPT-VQE has been primarily applied to molecular ground-state problems \cite{GrimsleyNatCommun2019, RamoanpjQuantumInf2025}, the algorithm has been implemented to several non-trivial problems, for example the preparation of the vacuum state of the Schwinger model on a 100-qubit superconducting quantum processor~\cite{FarrellPRXQuantum2024} and quantum dynamics simulations based on McLachlan’s variational principle~\cite{YaoPRXQuantum2021}. These applications illustrate the flexibility of the ADAPT-VQE framework and have motivated further developments in measurement reduction and alternative adaptive strategies~\cite{AnastasiouPRR2024,VaqueroSabaterJCTC2025, TancaraJCTC2026}.

In this work, we extend the ADAPT-VQE algorithm to qudit-based quantum computing. Specifically, we adopt a recent variant of ADAPT-VQE, where the operator pool is obtained from an approximate counterdiabatic Hamiltonian associated with an adiabatic evolution ending at the cost Hamiltonian ~\cite{TancaraJCTC2026} termed Counterdiabatic ADAPT-VQE (CD-ADAPT). This construction has been numerically shown for electronic structure problems that accelerates convergence toward the ground state while preserving the adaptive ansatz-building strategy of ADAPT-VQE. We formulate a ternary optimization problem in terms of angular-momentum operators, whose closed commutation relations make them particularly suitable for constructing approximate counterdiabatic Hamiltonian terms through nested-commutator expansions, and apply this formulation to the Max 3-Cut and Multiway Number Partitioning problems. We also investigate the the optimization landscape of this approach providing evidence that the proposed qudit-based adaptive construction retains the robustness against local minima observed in its qubit-based counterpart. The remainder of this work is organized as follows. Section~\ref{sec:methods} introduces the qudit Hamiltonians, the counterdiabatic construction of the operator pool and the adaptive algorithm; Section~\ref{sec:results} presents the numerical results for the Max 3-Cut and Multiway Number Partitioning problems, covering both the comparison against qudit QAOA and the analysis of local minima and barren plateaus; and Section~\ref{sec:conclusions} summarizes our conclusions. Technical details on the Hermiticity of the first-order pool, the compilation into native trapped-ion gates and the problem instances used are collected in Appendices~\ref{appendix1}--\ref{appendix_graphs}.

\section{Methods}
\label{sec:methods}
In this section, we present the mathematical framework used to define the cost and initial Hamiltonians in qudit-based systems. We then introduce the corresponding adiabatic evolution improved with counterdiabatic driving and its relation to quantum algorithms. Finally we describe the adaptive algorithm employed to construct the variational ansatz and the benchmark metrics considered in this work.

\subsection{Qudit operators for Hamiltonian encoding}
Qudit systems can be described using different operator formalisms, depending on both the physical encoding and the intended quantum algorithm. Among the most common choices are Weyl–Heisenberg operators, Gell-Mann matrices, transition operators between selected energy levels, and angular-momentum operators. Some of these representations can be directly related to the native controls available in specific experimental platforms. However, a universal qudit gate set has been experimentally demonstrated in trapped-ion processors \cite{RingbauerNatPhys2022}, allowing arbitrary multiqudit unitary operations to be synthesized from two-level equatorial rotations, acting on selected internal levels of a single ion, and generalized Mølmer–Sørensen gates, which generate entangling interactions between qudits. Consequently, algorithms formulated using different qudit operator representations can, in principle, be compiled into this universal native gate set. In this work, we adopt the angular-momentum representation to formulate the Hamiltonians for qudits (with $\hbar=1$):
\begin{align}
L_z \ket{\ell,m}
&=
m\ket{\ell,m},
\\
 L_\pm \ket{\ell,m} &=\sqrt{(\ell \mp m)(\ell\pm m+1)}\,
\ket{\ell,m\pm1},
\\
L_x
&=
\frac{1}{2}\left(L_+ + L_-\right), \qquad L_y
=
\frac{1}{2i}\left(L_+ - L_-\right),
\end{align}
where the angular-momentum operators satisfy the commutator properties:
\begin{equation}
[L_i,L_j] = i\,\varepsilon_{ijk}L_k
\end{equation}
where $i,j,k\in\{x,y,z\}$ and $\varepsilon_{ijk}$ is the Levi-Civita symbol. In this work, we focus on optimization problems involving ternary variables. Therefore, it is sufficient to consider the \mbox{$\ell=1$} representation, corresponding to a three-dimensional local Hilbert space, for which the angular-momentum operators are given by:
\begin{equation}
\begin{aligned}
L_x &= \frac{1}{\sqrt{2}}
\begin{pmatrix}
0 & 1 & 0 \\
1 & 0 & 1 \\
0 & 1 & 0
\end{pmatrix}, \\
L_y &= \frac{1}{\sqrt{2}}
\begin{pmatrix}
0 & -i & 0 \\
i & 0 & -i \\
0 & i & 0
\end{pmatrix}, \\
L_z &= 
\begin{pmatrix}
1 & 0 & 0 \\
0 & 0 & 0 \\
0 & 0 & -1
\end{pmatrix}.
\end{aligned}
\end{equation}
The operator $L_z$ can be used to represent a ternary variable $v_i \in \{-1,0,1\}$, since these three values correspond to its eigenvalues and $L_z$ is diagonal in the computational basis:
\begin{equation}
|{-1}\rangle =
\begin{pmatrix}
0\\
0\\
1
\end{pmatrix},
\qquad
|0\rangle =
\begin{pmatrix}
0\\
1\\
0
\end{pmatrix},
\qquad
|1\rangle =
\begin{pmatrix}
1\\
0\\
0
\end{pmatrix}.
\label{eq:basis}
\end{equation}

Since we consider discrete classical optimization problems involving ternary variables, the cost Hamiltonian encoding the optimization problem can be written as:
\begin{equation}
H_C =
\sum_j \omega_j^{(1)} L_z^{(j)}
+
\sum_{j,k} \omega_{j,k}^{(2)} L_z^{(j)}L_z^{(k)}
+
\sum_{j,k,l} \omega_{j,k,l}^{(3)}
L_z^{(j)}L_z^{(k)}L_z^{(l)}
+\cdots ,
\end{equation}
 where $L_z^{(j)}=I^{\otimes (j-1)}\otimes L_z\otimes I^{\otimes (n-j)}$ denotes the operator $L_z$ acting on the $j$-th of the $n$ qutrits, and the real coefficients $\omega_j^{(1)}$, $\omega_{j,k}^{(2)}$, $\omega_{j,k,l}^{(3)},\ldots$ encode the classical cost function of the problem.
 
The objective is to minimize the expectation value of $H_C$, or equivalently, to find its ground state. Since $H_C$ is diagonal in the computational basis, its ground state represents a configuration of ternary variables that minimizes the encoded cost function and therefore corresponds to an optimal solution to the original optimization problem. In quantum algorithms for optimization problems, it is common to introduce an additional Hamiltonian known as the mixer Hamiltonian. Inspired by adiabatic quantum evolution, which will be reviewed in the next section, the mixer Hamiltonian is chosen such that its ground state is an equal-weight superposition of all computational-basis states. In this work we use the following mixer Hamiltonian:

\begin{equation}
H_M = -\sum_{j=1}^{n}
\left[
\sqrt{2}\,L_x^{(j)}
+
\left(L_z^{(j)}\right)^2
\right].\label{mixer-qudit-adapt}
\end{equation}
The specific combination $-\left(\sqrt{2}\,L_x+L_z^2\right)$ is chosen because the uniform superposition state: 
\begin{equation}
\left|+_3\right\rangle
=
\frac{1}{\sqrt{3}}
\left(
\left|-1\right\rangle
+
\left|0\right\rangle
+
\left|1\right\rangle
\right)
\end{equation}
is its ground state. Since $H_M$ is a sum of independent single-qutrit terms, the ground state $\left|\phi_g\right\rangle$ of $H_M$ for a system of $n$ qutrits is the product state
\begin{equation}
\left|\phi_g\right\rangle
=
\bigotimes_{j=1}^{n} \left|+_3\right\rangle_j,
\end{equation}
which is the uniform superposition of all $3^n$ computational-basis states.

\subsection{Counterdiabatic improvement of adiabatic evolution}
Adiabatic quantum computing exploits the adiabatic theorem to prepare the ground state of a cost Hamiltonian $H_C$, starting from a known ground state of an initial Hamiltonian, in this case the mixer Hamiltonian $H_M$. The evolution is generated by the adiabatic Hamiltonian:
\begin{equation}
H_{\mathrm{ad}}(t)
=
\left[1-\lambda(t)\right]H_M
+
\lambda(t)H_C,
\end{equation}
where the schedule function satisfies the boundary conditions $\lambda(0)=0$ and $\lambda(T)=1$, with $T$ denoting the total evolution time. Therefore, the adiabatic Hamiltonian starts from $H_M$ at $t=0$ and evolves continuously toward the cost Hamiltonian $H_C$ at $t=T$. The state evolved under the adiabatic Hamiltonian is given by:
\begin{equation}
\ket{\psi(T)}
=
\mathcal{T}
\exp\left[
-i\int_{0}^{T}
H_{\mathrm{ad}}(t)\,dt
\right]
\ket{\psi(0)},
\end{equation}
where $\mathcal{T}$ denotes the time-ordering operator. If we begin with $\ket{\psi(0)}=\ket{\phi_g}$, the ground state of $H_M$, therefore at the final time $T$ we end in the ground state of $H_C$ obtaining the solution of the optimization problem. This adiabatic evolution motivates QAOA \cite{FarhiQAOA2014}, whose variational ansatz can be interpreted as a digitized approximation to the adiabatic evolution that generates $\ket{\psi(T)}$. The ansatz in QAOA is given by:

\begin{equation}
\ket{\psi_p(\boldsymbol{\gamma},\boldsymbol{\beta})}
=
\prod_{k=1}^{p}
e^{-i\beta_k H_M}
e^{-i\gamma_k H_C}
\ket{\phi_g},
\end{equation}
where $\boldsymbol{\beta}=(\beta_1,\ldots,\beta_p)$ and $\boldsymbol{\gamma}=(\gamma_1,\ldots,\gamma_p)$ are vectors of variational parameters, optimized to approximate the ground state of $H_C$ by minimizing the cost function
\begin{equation}
E_{\mathrm{QAOA}}
= \min_{(\boldsymbol{\gamma},\boldsymbol{\beta})}
\bra{\psi_p(\boldsymbol{\gamma},\boldsymbol{\beta})}
H_C
\ket{\psi_p(\boldsymbol{\gamma},\boldsymbol{\beta})}.\label{E_QAOA}
\end{equation}
QAOA has also been extended to qudit-based systems, where analogous qudit operators are used to encode the mixer and cost Hamiltonians \cite{DellerPRA2023}.

However the adiabatic theorem requires sufficiently slow evolution and direct implementation of the adiabatic protocol may demand prohibitively long evolution times. This limitation can be mitigated through counterdiabatic driving, which introduces an additional term designed to suppress diabatic transitions and accelerate the preparation of the target state. The resulting Hamiltonian is given by
\begin{equation}
H_{\mathrm{cd}}(t)
=
\left[1-\lambda(t)\right]H_M
+
\lambda(t)H_C
+
\dot{\lambda}(t)A_{\lambda},
\end{equation}
where $\dot{\lambda}(0)=\dot{\lambda}(T)=0$ and $A_{\lambda}$ is the adiabatic gauge potential (AGP). The exact AGP generally depends on the instantaneous eigenstates and eigenvalues of the adiabatic Hamiltonian, making its direct evaluation impractical for many-body systems. Nevertheless, approximate AGPs can be constructed without explicitly diagonalizing $H_{\mathrm{ad}}(t)$. In particular, the nested-commutator expansion introduced in Ref.~\cite{ClaeysPRL2019} expresses the AGP at order $l$ as
\begin{equation}
A_{\lambda}^{(l)}
=
i\sum_{q=1}^{l}
\alpha_q(t)O_{2q-1}(t), \label{AGP}
\end{equation}
where the operators are recursively defined as
\begin{equation}
O_0(t)=\partial_{\lambda}H_{\mathrm{ad}}(t),
\qquad
O_k(t)=
\left[
H_{\mathrm{ad}}(t),
O_{k-1}(t)
\right],
\end{equation}
and the coefficients $\alpha_q(t)$ can be obtained by minimizing the action $S_l = \Tr{[G_l^2]}$, with $G_l =\partial_\lambda \hat{H}_{\textrm{ad}}- i[\hat{H}_{\text{ad}}, \hat{A}_{\lambda}^{(l)} ]$.
As the expansion order increases, the approximate AGP systematically approaches the exact AGP, recovering it in the limit $l\rightarrow\infty$.

Counterdiabatic driving with approximate AGP have been applied to qubit-based quantum computing, commonly using the first-order expansion $l=1$ of nested commutators \cite{HegadePRResearch2022}. The evolution generated by $H_{\mathrm{cd}}(t)$ can also be used to construct a variational ansatz, leading to the digitized counterdiabatic QAOA \cite{ChandaranaPRR2022}.
These approaches have shown improvements in state-preparation performance and circuit resources compared with conventional adiabatic approaches \cite{HegadePRResearch2022, ChandaranaPRR2022} and the extension to qudit-based quantum computing was recently proposed \cite{TancaraNPJQuantumInf2025, BottarelliPRR2025}.

\subsection{CD-ADAPT for qudit-based quantum computing}

The variational quantum eigensolver (VQE) approximates the ground state of a cost Hamiltonian by preparing a parametrized state $|\psi(\boldsymbol{\theta})\rangle$ and minimizing its energy through a classical optimization procedure. The variational energy is defined as

\begin{equation}
E_{\mathrm{VQE}}
=
\min_{\boldsymbol{\theta}}
\langle \psi(\boldsymbol{\theta}) |
H_C
| \psi(\boldsymbol{\theta}) \rangle .
\label{eq:vqe_energy}
\end{equation}
The VQE cost function in Eq.~\eqref{eq:vqe_energy} has the same structure as the QAOA cost function introduced in Eq.~\eqref{E_QAOA}. In fact, QAOA can be regarded as a VQE with a fixed ansatz inspired by adiabatic evolution. Fixed ansatz constructions must balance expressibility and circuit depth, since an insufficiently expressive ansatz may fail to approximate the target state, whereas excessively deep and expressive circuits can lead to a concentration of the cost function and exponentially vanishing gradients, giving rise to barren plateaus \cite{LaroccaNatRevPhys2025}. ADAPT-VQE addresses these limitations by iteratively constructing compact, problem-dependent ansatzes using an energy-gradient criterion over a predefined operator pool, while showing improved robustness against barren plateaus \cite{GrimsleynpjQuantumInf2023}. 

More recently, CD-ADAPT \cite{TancaraJCTC2026} was introduced as an adaptive variational algorithm in which the operator pool is derived from counterdiabatic driving through nested-commutator approximations to the AGP, thereby incorporating information from the adiabatic interpolation into the adaptive ansatz construction. This choice is physically motivated: the AGP suppresses diabatic transitions and identifies directions in Hilbert space that are relevant for reaching the ground state of $H_C$. This approach has shown higher accuracy and lower circuit depth than ADAPT-VQE with standard operator pools in molecular simulations. The operator pool is constructed by considering the impulse regime \cite{CadavidPhysRevApplied2024} where the dynamics is mainly governed by the rate of change of the schedule function, under the condition $|\lambda(t)| \ll |\dot{\lambda}(t)|$. Consequently, the counterdiabatic Hamiltonian can be approximated as

\begin{equation}
H_{\mathrm{cd}}(t)
\approx
\dot{\lambda}(t) A_{\lambda}^{(l)}.
\label{eq:cd_hamiltonian}
\end{equation}

For qudit systems the $l$-th order approximation to the AGP obtained with nested-commutators can be expressed as a linear combination of angular-momentum operator strings:

\begin{equation}
A_{\lambda}^{(l)}
=
\sum_{j=1}^{\eta}
a_j(t) D_j,
\label{eq:agp_generator_expansion}
\end{equation}
where $\eta$ denotes the number of angular-momentum operator strings appearing in $A_{\lambda}^{(l)}$, while $a_j(t)$ is the weight associated with the angular-momentum operator string $D_j$ obtained from the corresponding $\alpha_q(t)$ coefficients and from combinations of powers of $\lambda(t)$ and $1-\lambda(t)$ arising from the nested-commutator expansion. Here, an important difference from the qubit case arises. For qubits, the operators $\{D_j\}_{j=1}^{\eta}$ can be used directly to construct the operator pool because they are Pauli strings, and every Pauli string is Hermitian. In the qudit case, however, quadratic angular-momentum operators (such as $L_z^{2}$ appearing in the mixer or cost Hamiltonian) can generate products of noncommuting angular-momentum components acting on the same qudit that are not necessarily Hermitian when treated as individual strings. For example, consider the commutator generated by the mixer term
$L_x^{(i)}$ and the two-local cost term
$L_z^{(i)}L_z^{(j)}$:
\begin{equation}
\left[
L_x^{(i)},
L_z^{(i)}L_z^{(j)}
\right]
=
-iL_y^{(i)}L_z^{(j)}.
\end{equation}
A subsequent commutator with the quadratic mixer term
$\left(L_z^{(i)}\right)^2$ gives
\begin{equation}
\begin{aligned}
&
\left[
\left(L_z^{(i)}\right)^2,
L_y^{(i)}L_z^{(j)}
\right] =
-i\left(
L_z^{(i)}L_x^{(i)}L_z^{(j)}
+
L_x^{(i)}L_z^{(i)}L_z^{(j)}
\right).
\end{aligned}
\end{equation}
We can see that the angular-momentum operator string $L_z^{(i)}L_x^{(i)}L_z^{(j)}$ appears, containing the non-Hermitian product $L_zL_x$ acting on qudit $i$. Therefore, in general, the set $\{D_j\}_{j=1}^{\eta}$ may contain non-Hermitian operators. To address this issue, we define the following new set of operators:
\begin{equation}
\mathcal{V}
=
\left\{V_j\right\}_{j=1}^{\eta'},
\qquad
V_j
=
\frac{D_j+D_j^\dagger}{2},
\end{equation}
where $\eta'$ is the number of operators obtained after removing symbolically identical expressions. The resulting set $\mathcal{V}$ is used as the operator pool for CD-ADAPT because its operators retain their connection to the approximate AGP, as discussed in Appendix~\ref{appendix1}. The next step is the construction of the ansatz from the operator pool $\mathcal{V}$, following the standard ADAPT-VQE algorithm. The algorithm is initialized at iteration $k=0$ with the state $\left|\phi_g\right\rangle$ because it is the ground state of $H_M$. At each iteration $k$, the energy gradient associated with
each operator $V_j$ is evaluated at $\theta_j=0$ as
\begin{equation}
g_j^{(k)}
=
\left.
\frac{\partial E}{\partial\theta_j}
\right|_{\theta_j=0}
=
i\langle\psi_k|[V_j,H_C]|\psi_k\rangle .
\label{eq:adapt_gradient}
\end{equation}

The collection of these gradients defines the gradient vector
$\boldsymbol{g}^{(k)}=(g_1^{(k)},\ldots,g_{\eta'}^{(k)})$. If
$\|\boldsymbol{g}^{(k)}\|_2<\varepsilon$, the algorithm terminates, where $\varepsilon$ is a predefined convergence threshold. Otherwise,
the operator associated with the largest gradient magnitude is selected,
\begin{equation}
j^{*}
=
\underset{j}{\operatorname{arg\,max}}
\left|g_j^{(k)}\right|,
\label{eq:adapt_selection}
\end{equation}
and the corresponding unitary is added to the ansatz,
\begin{equation}
|\psi_{k+1}\rangle
=
e^{-i\theta_{k+1}V_{j^{*}}}
|\psi_k\rangle .
\label{eq:adapt_state_update}
\end{equation}

A VQE optimization is then performed over all accumulated variational
parameters. Denoting by $\boldsymbol{\theta}_{k}^{*}$ the optimal parameters obtained at iteration $k$, the optimization is initialized at $\boldsymbol{\theta}_{k+1} = (\boldsymbol{\theta}_{k}^{*}, 0)$, that is, the previously optimized parameters are reused and the newly introduced parameter is set to zero. Thus, the algorithm follows a warm-start strategy. The resulting optimized state $|\psi_{k+1}^{\mathrm{opt}}\rangle$ is used as
the input state for the next iteration. This procedure is repeated until
$\|\boldsymbol{g}^{(k)}\|_2<\varepsilon$.

As can be seen from the gradient expression in Eq. \ref{eq:adapt_gradient} and the nested-commutator
expansion of the approximate AGP in Eq. \ref{AGP}, the construction naturally favors an
operator basis whose elements obey closed commutation relations. In this respect, angular-momentum operators provide a convenient representation, which motivates our choice of these operators for qudits. It is important to note that the number $\eta$ of operators generated by the nested commutators can increase considerably with the truncation order $l$ of the approximate AGP, as observed in the original CD-ADAPT proposal \cite{TancaraJCTC2026}. However, this does not directly translate into a larger circuit depth, because the generated operators are not all included in the ansatz. Instead, only the size of the operator pool increases. Moreover, although $\eta$ increases with $l$, it does not grow exponentially and remains a very small fraction of the total number of possible operator strings as observed in \cite{TancaraJCTC2026}.

Finally, because the construction is based on an adiabatic interpolation between an initial and a final Hamiltonian, the CD-ADAPT framework is, in principle, applicable to a broad class of problems whose Hamiltonians can be represented in terms of angular-momentum operators and for which the corresponding approximate AGP
can be constructed. In the following, we refer to the CD-ADAPT algorithm employing the operator pool $\mathcal{V}$ for qudit-based optimization problems as Qudit-ADAPT.

\onecolumngrid
\begin{center}
    \includegraphics[width=0.9\textwidth]{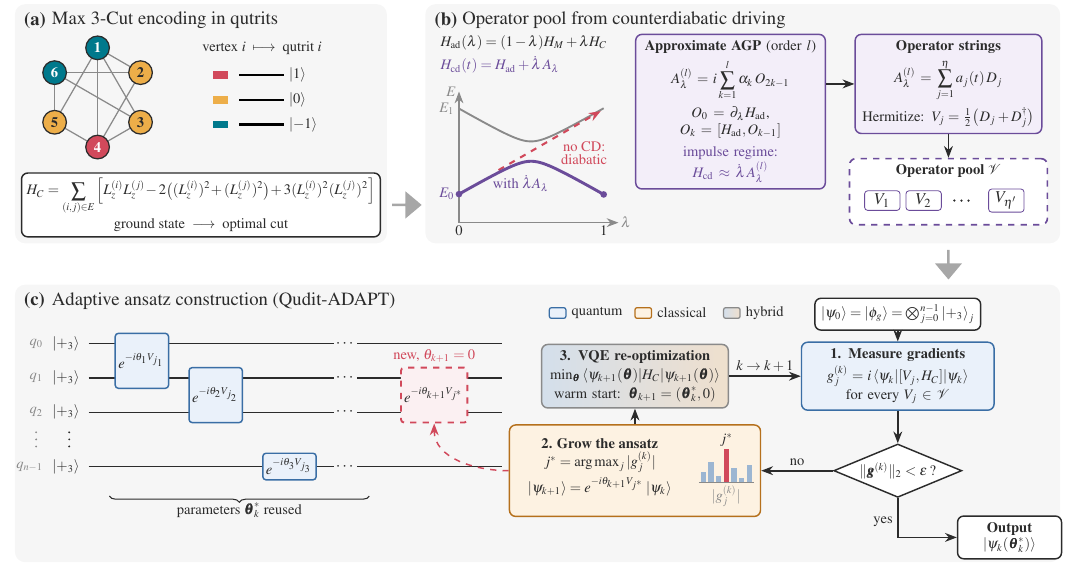}
     \captionof{figure}{Schematic illustration of the Qudit-ADAPT algorithm.
(a) Example of a combinatorial optimization problem, Max 3-Cut, encoded using qutrits, where each graph vertex is associated with a qutrit whose computational basis states represent the three possible partitions.
(b) Construction of the operator pool from an approximate AGP, obtained through a nested-commutator expansion of the adiabatic Hamiltonian. The resulting angular-momentum operator strings are Hermitized to form the pool of candidate generators.
(c) Adaptive construction of the variational ansatz. At each iteration, the energy gradients are evaluated for all operators in the pool, and the operator with the largest gradient magnitude is added to the circuit. All variational parameters are then reoptimized using a warm-start strategy. The procedure is repeated until the gradient norm falls below a predefined convergence threshold.}\label{qudit-adapt}
\end{center}
 \twocolumngrid
\section{Results}
\label{sec:results}
In this section, we present numerical simulation results for Qudit-ADAPT by considering optimization problems that admit a formulation in terms of ternary variables. We use several metrics to quantify the performance and compare \mbox{Qudit-ADAPT}
with qudit QAOA as an algorithm with a fixed ansatz.  While Qudit-ADAPT employs the mixer Hamiltonian defined in Eq.~\ref{mixer-qudit-adapt}, QAOA uses a mixer consisting of a sum of local $L_x$ operators acting on individual qutrits, given by $H_M^{\mathrm{QAOA}}=\sum_{i=1}^{n}L_x^{(i)}$ as originally proposed in Ref. \cite{DellerPRA2023}. We use \texttt{QuTiP} \cite{Lambert2026} to construct and manipulate the angular-momentum operators, while the VQE parameters are optimized using the BFGS method implemented in \texttt{SciPy} \cite{Virtanen2020}.

\subsection{Max 3-Cut}
We consider the Max 3-Cut problem, which is the particular case of the Max $k$-Cut problem for $k=3$. Given a graph $G=(V,E)$, the Max $k$-Cut problem consists of partitioning the vertex set $V$ into $k$ disjoint subsets such that the number of edges connecting vertices assigned to different subsets is maximized. For the case $k=3$, the problem admits a natural formulation in terms of ternary variables, where each vertex is assigned one ternary variable $v_i \in \{-1,0,1\}$. Each value identifies one of the three subsets, with $-1$, $0$,
and $1$ corresponding to the first, second, and third subsets, respectively. This problem has previously been addressed in the context of qudit-based quantum computing \cite{TancaraNPJQuantumInf2025, BottarelliPRR2025}, with the corresponding cost Hamiltonian given by:
\begin{equation}
H_C
=
\sum_{\{i,j\}\in E}
\left[
L_z^{(i)}L_z^{(j)}
-2\left(
\left(L_z^{(i)}\right)^2
+
\left(L_z^{(j)}\right)^2
\right)
+3\left(L_z^{(i)}\right)^2
\left(L_z^{(j)}\right)^2
\right].
\label{eq:maxcut_hc}
\end{equation}
\onecolumngrid
\begin{center}
    \vspace{0.8em}
    \begin{minipage}{0.32\textwidth}
        \centering
        \begin{overpic}[width=0.99\linewidth]{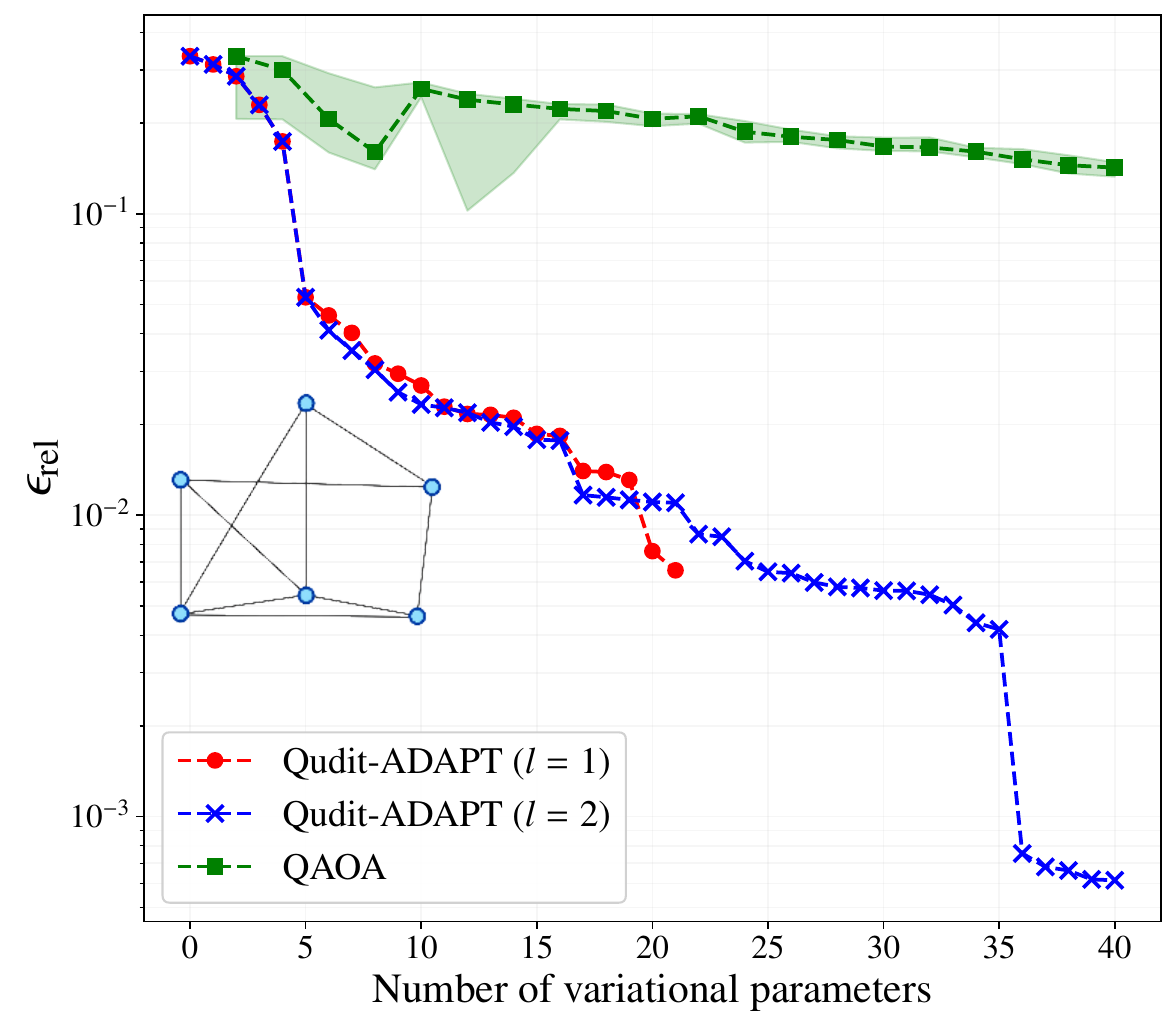}
            \put(2,88){a)}
        \end{overpic}

        \vspace{0.4em}

        \begin{overpic}[width=0.99\linewidth]{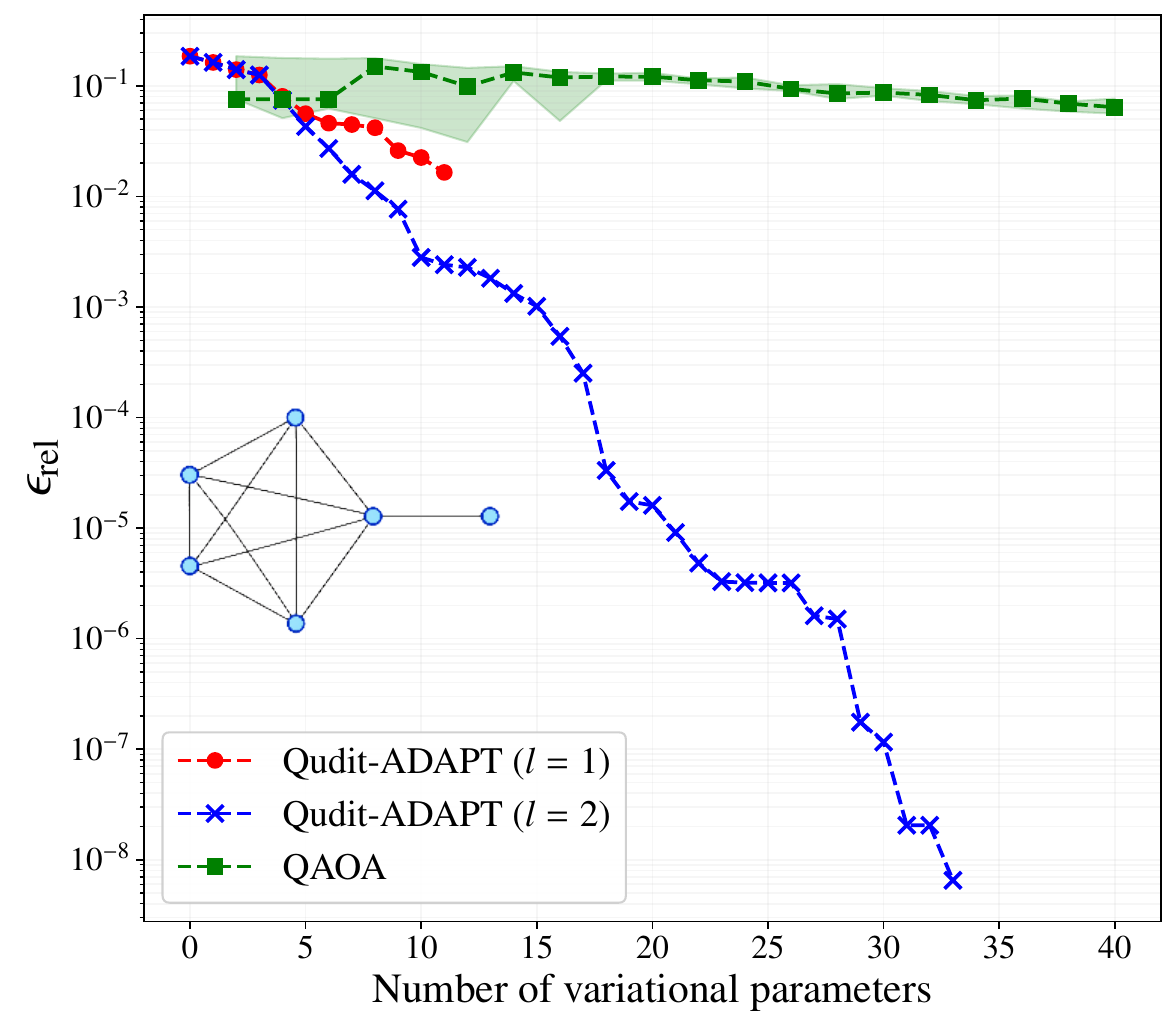}
            \put(2,88){d)}
        \end{overpic}
    \end{minipage}%
    \hfill
    \begin{minipage}{0.32\textwidth}
        \centering
        \begin{overpic}[width=0.99\linewidth]{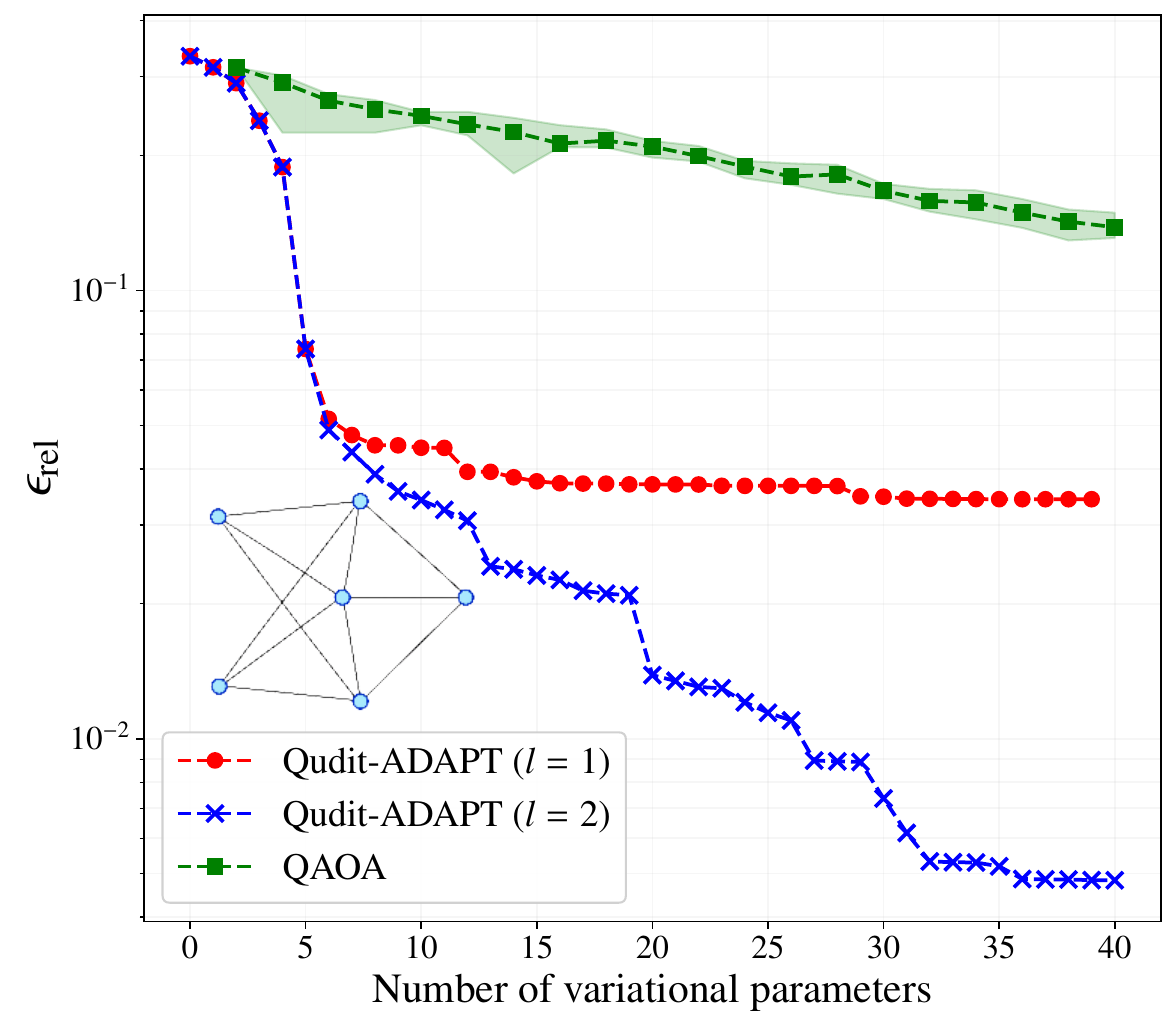}
            \put(2,88){b)}
        \end{overpic}

        \vspace{0.4em}

        \begin{overpic}[width=0.99\linewidth]{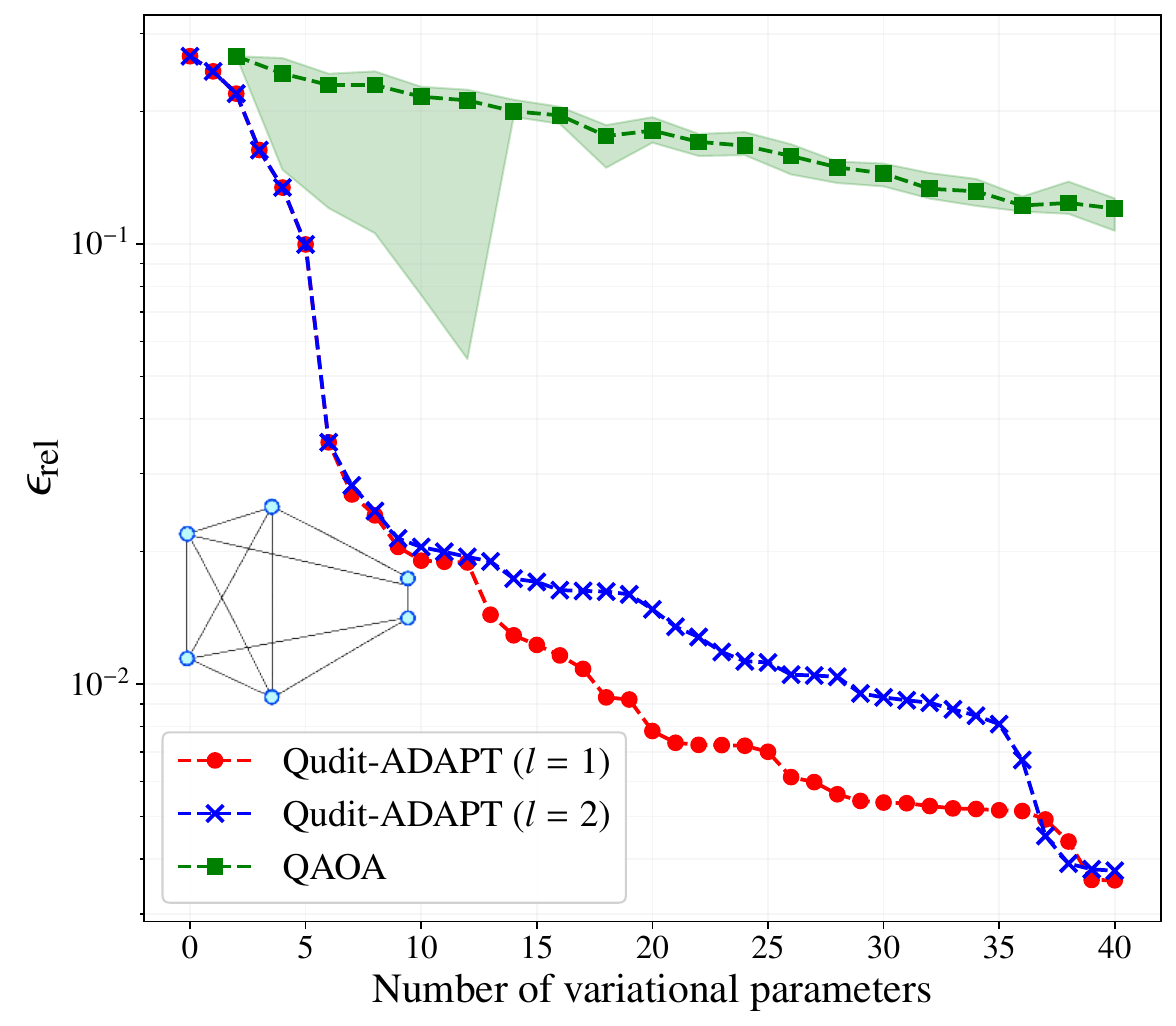}
            \put(2,88){e)}
        \end{overpic}
    \end{minipage}%
    \hfill
    \begin{minipage}{0.32\textwidth}
        \centering
        \begin{overpic}[width=0.99\linewidth]{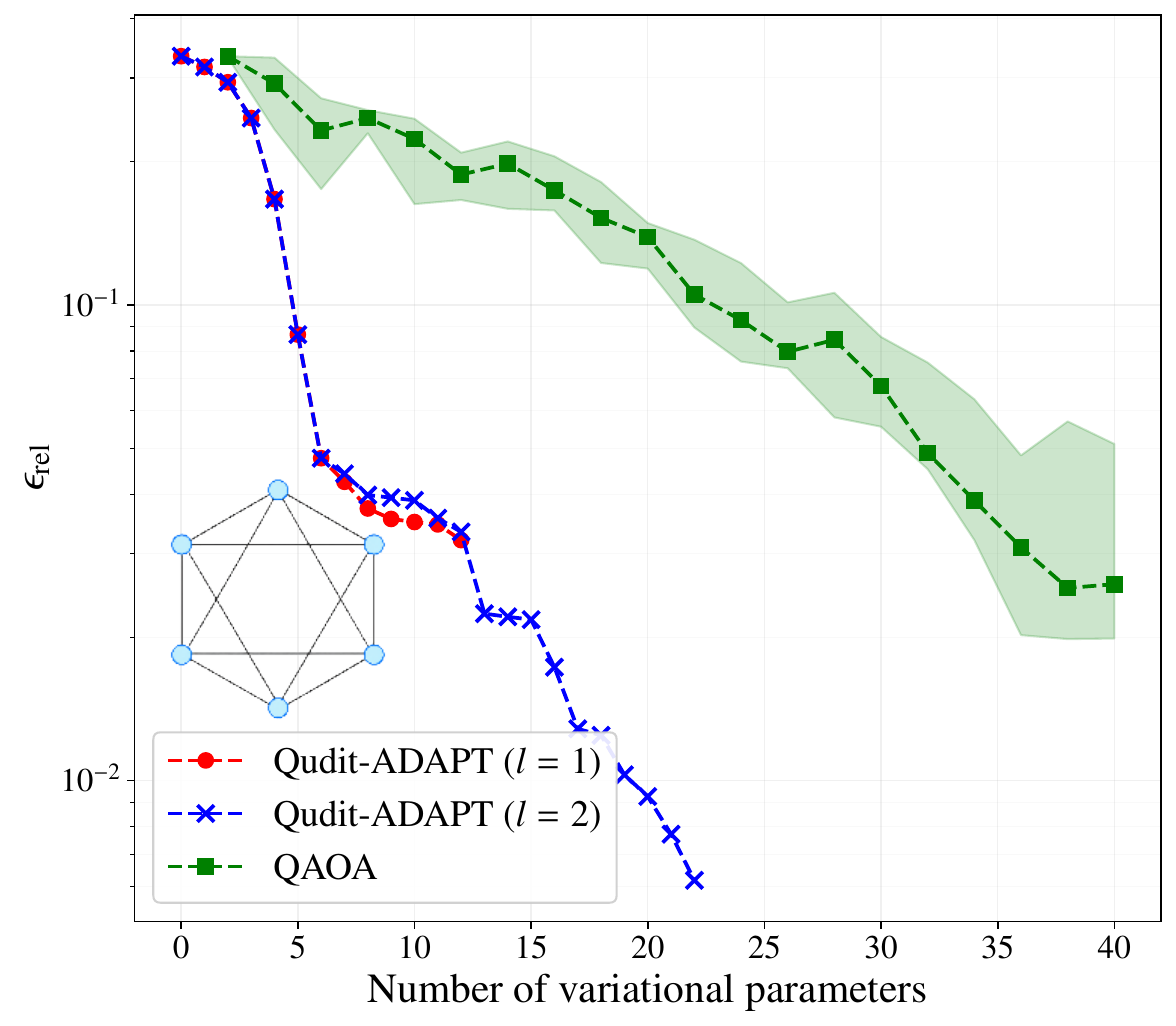}
            \put(2,88){c)}
        \end{overpic}

        \vspace{0.4em}

        \begin{overpic}[width=0.99\linewidth]{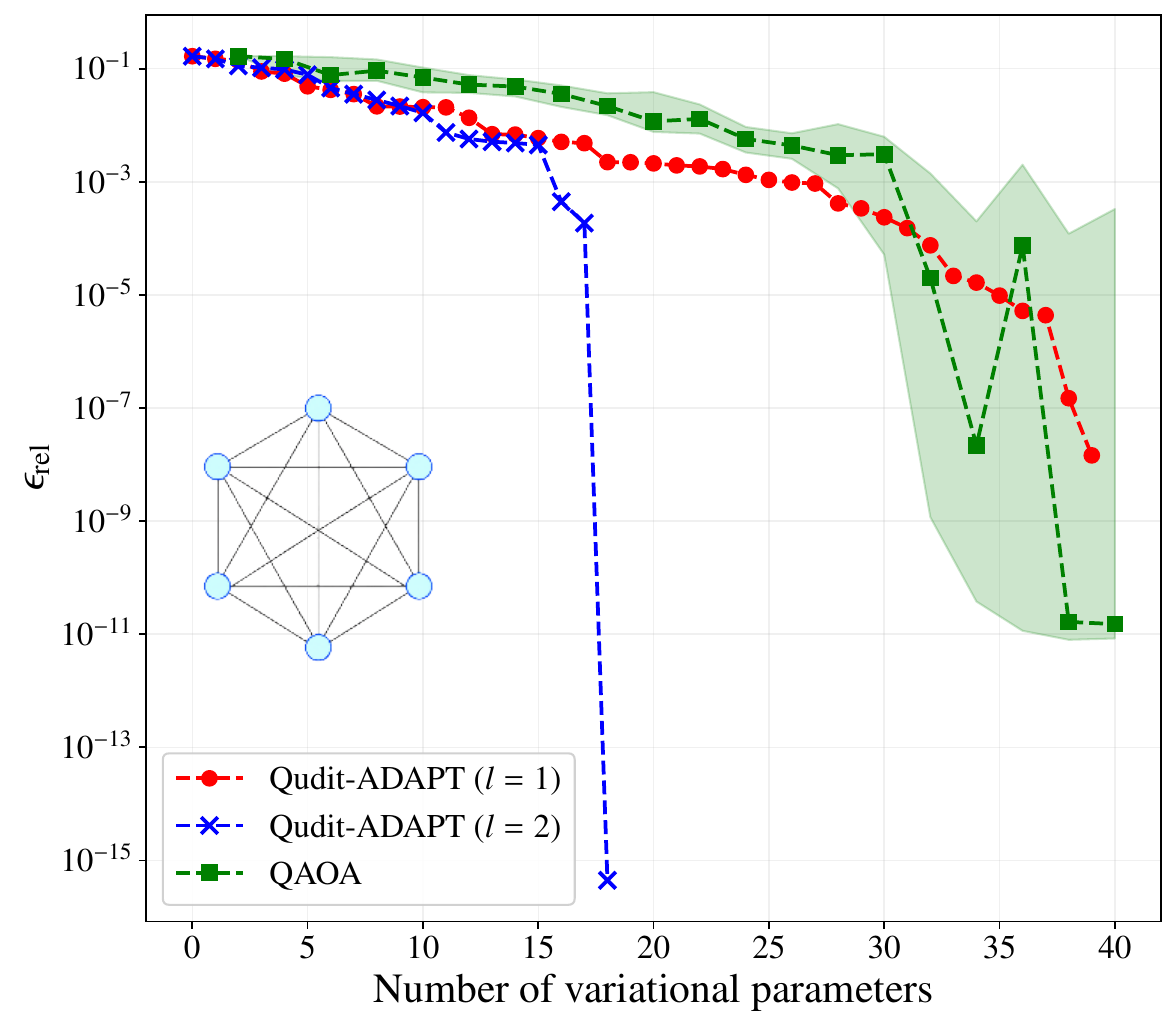}
            \put(2,88){f)}
        \end{overpic}
    \end{minipage}

    \captionof{figure}{Relative error as a function of the number of variational parameters for Qudit-ADAPT using operator pools obtained from the $l=1$ and $l=2$ truncations of the approximate AGP, and for QAOA using 25 random restarts. For QAOA, the median is plotted, while the shaded region represents the interquartile range. a) $G_1$, b) $G_3$, c) $G_5$, d) $G_2$, e) $G_4$, and f) $K_6$. Instances $G_1$--$G_4$ are irregular graphs, while $G_5$ and $K_6$ are regular (degree 4, and degree 5 complete, respectively). The graph corresponding to each instance is shown as an inset.}
    \label{fig_instances}
\end{center}
\twocolumngrid

We first solve this problem for four irregular graph instances with six vertices, differing in their connectivity and ground-state degeneracy. Six-vertex instances provide nontrivial irregular graphs while remaining small enough to allow exact classical simulation of the full Hilbert space, without resorting to approximate methods. Details of the graph connectivity are provided in Appendix~\ref{appendix_graphs}. The metric considered here is the relative error:
\begin{equation}
\epsilon_{\mathrm{rel}}
=
\frac{\left|E_{\mathrm{final}}-E_{0}\right|}
{\left|E_{0}\right|},
\label{eq:relative_error}
\end{equation}
where $E_{0}$ is the exact ground-state energy obtained by diagonalizing $H_{C}$, and
\begin{equation}
E_{\mathrm{final}}
=
\bra{\psi(\boldsymbol{\theta}^{*})}
H_{C}
\ket{\psi(\boldsymbol{\theta}^{*})}
\end{equation}
is the energy obtained by the algorithm, with $\ket{\psi(\boldsymbol{\theta}^{*})}$ denoting the state evaluated at the optimized parameters $\boldsymbol{\theta}^{*}$.

Figure~\ref{fig_instances} compares the performance of Qudit-ADAPT and QAOA as a function of the number of variational parameters. In Qudit-ADAPT, the number of parameters increases adaptively during the construction of the ansatz. At each iteration, the optimized parameters obtained in the previous VQE step are used as a warm-start, and the operator with the largest gradient magnitude is selected from the operator pool and appended to the ansatz until convergence, where we considered the threshold $\varepsilon = 10^{-4}$ and a maximum of 40 operators. By contrast, QAOA employs a fixed ansatz without warm-start initialization. 

We consider up to $p=20$ QAOA layers, corresponding to 40 variational parameters, since each layer introduces two parameters: one associated with the cost Hamiltonian and another with the mixer Hamiltonian. For each value of $p$, the optimization is performed using 25 random parameter initializations, and the median relative error is reported. For the four instances considered, Qudit-ADAPT with the $l=2$ pool achieves a relative error at least one order of magnitude smaller than the median obtained with QAOA at $p=20$, while with the $l=1$ pool the reduction ranges from a factor of about four ($G_2$ and $G_3$) to more than one order of magnitude ($G_1$ and $G_4$), as summarized in Table~\ref{table1}. This improvement is consistent with the combined effect of the warm-start strategy and the adaptive construction of the ansatz, in which operators are selected according to a gradient criterion from pools derived using counterdiabatic driving.

We also consider two regular graphs, shown in panels c) and f) of Fig.~\ref{fig_instances}. In these more symmetric instances, QAOA exhibits improved performance, while Qudit-ADAPT still reaches solutions of comparable accuracy. Notably, for the degree-5 complete graph $K_6$ and regular graph $G_5$, QAOA outperforms Qudit-ADAPT with the $l=1$ pool, although it exhibits a larger interquartile range. With the $l=2$ pool, however, Qudit-ADAPT outperforms QAOA, reaching machine precision after only 18 parameters for $K_6$, well beyond the accuracy attained by QAOA at $p=20$; the additional operators available at $l=2$ are therefore what makes the difference on this highly connected instance.

We also consider the approximation ratio $r$ defined as:
\begin{equation}
    r=\left|
\frac{E_{\mathrm{final}}}{E_0}
\right|,
\end{equation}
\begin{table}[H]
\centering
\begin{tabular}{c c|cc|cc}
\hline\hline
& \multicolumn{1}{c|}{QAOA ($p = 20$)}
& \multicolumn{2}{c|}{Qudit-ADAPT $l = 1$}
& \multicolumn{2}{c}{Qudit-ADAPT $l = 2$} \\
Instance & $r_{\mathrm{med}}$ & $r$ & Params. & $r$ & Params. \\
\hline
$G_1$ & 0.858 & 0.993 & 21 & 0.999 & 40 \\
$G_2$ & 0.936 & 0.984 & 11 & $\approx 1.000$ & 33 \\
$G_3$ & 0.855 & 0.966 & 39 & 0.995 & 40 \\
$G_4$ & 0.891 & 0.996 & 40 & 0.996 & 40 \\
$G_5$ & 0.971 & 0.968 & 12 & 0.994 & 22 \\
$K_6$ & $\approx 1.000$ & $\approx 1.000$ & 39 & $\approx 1.000$ & 18 \\

\hline\hline
\end{tabular}
\caption{Comparison of QAOA and Qudit-ADAPT for the benchmark instances $G_1$--$G_4$ (irregular graphs) and the regular graphs $G_5$ (degree 4) and $K_6$ (degree 5, complete), all shown in Fig.~\ref{fig_instances}.}
\label{table1}
\end{table}
Table \ref{table1} summarizes the results of these particular instances, where $r_{\mathrm{med}}$ for QAOA is considering the 25 random parameter initializations and with $p=20$. We observe that Qudit-ADAPT generally achieves higher approximation ratios than QAOA and, in several cases, requires fewer variational parameters. In addition, QAOA exhibits greater variability across random initializations, whereas the warm-start strategy used in Qudit-ADAPT leads to more consistent convergence. It should also be noted that, in most cases, each Qudit-ADAPT parameter is associated with the exponential of a single angular-momentum operator string, while each QAOA parameter multiplies a complete Hamiltonian composed of multiple operator strings. Moreover, a comparison in terms of native-gate decompositions confirms that the Qudit-ADAPT ansatz has a lower implementation cost. Further details are provided in Appendix~\ref{app:gate_cost}.  These results indicate that Qudit-ADAPT can achieve higher accuracy with lower variational and circuit-resource requirements.
\begin{figure}[H]
    \centering

    \begin{subfigure}
        \centering
        \begin{overpic}[width=0.95\linewidth]{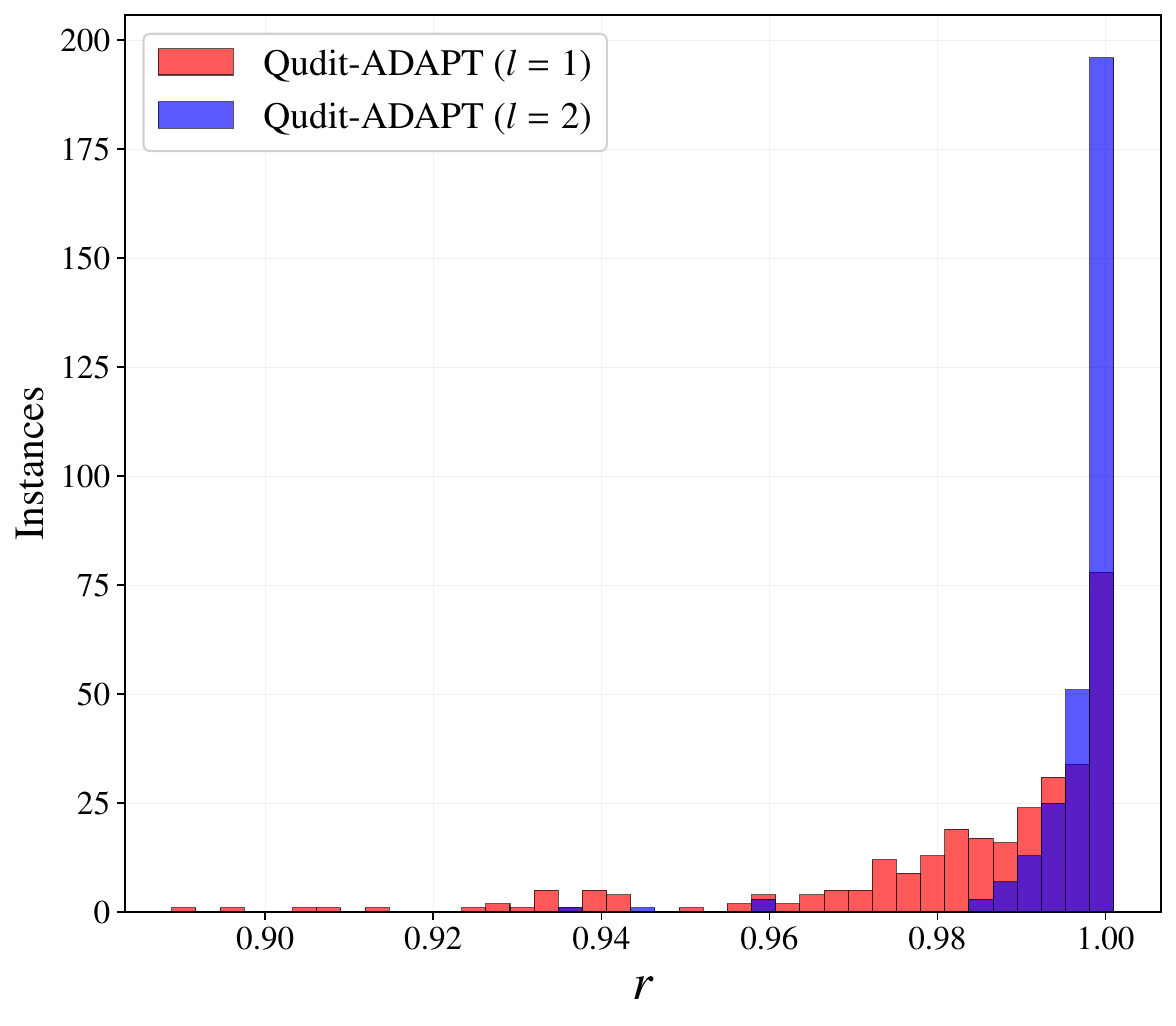}
            \put(2,88){a)}
        \end{overpic}
    \end{subfigure}

    \vspace{0.4em}

    \begin{subfigure}
        \centering
        \begin{overpic}[width=0.95\linewidth]{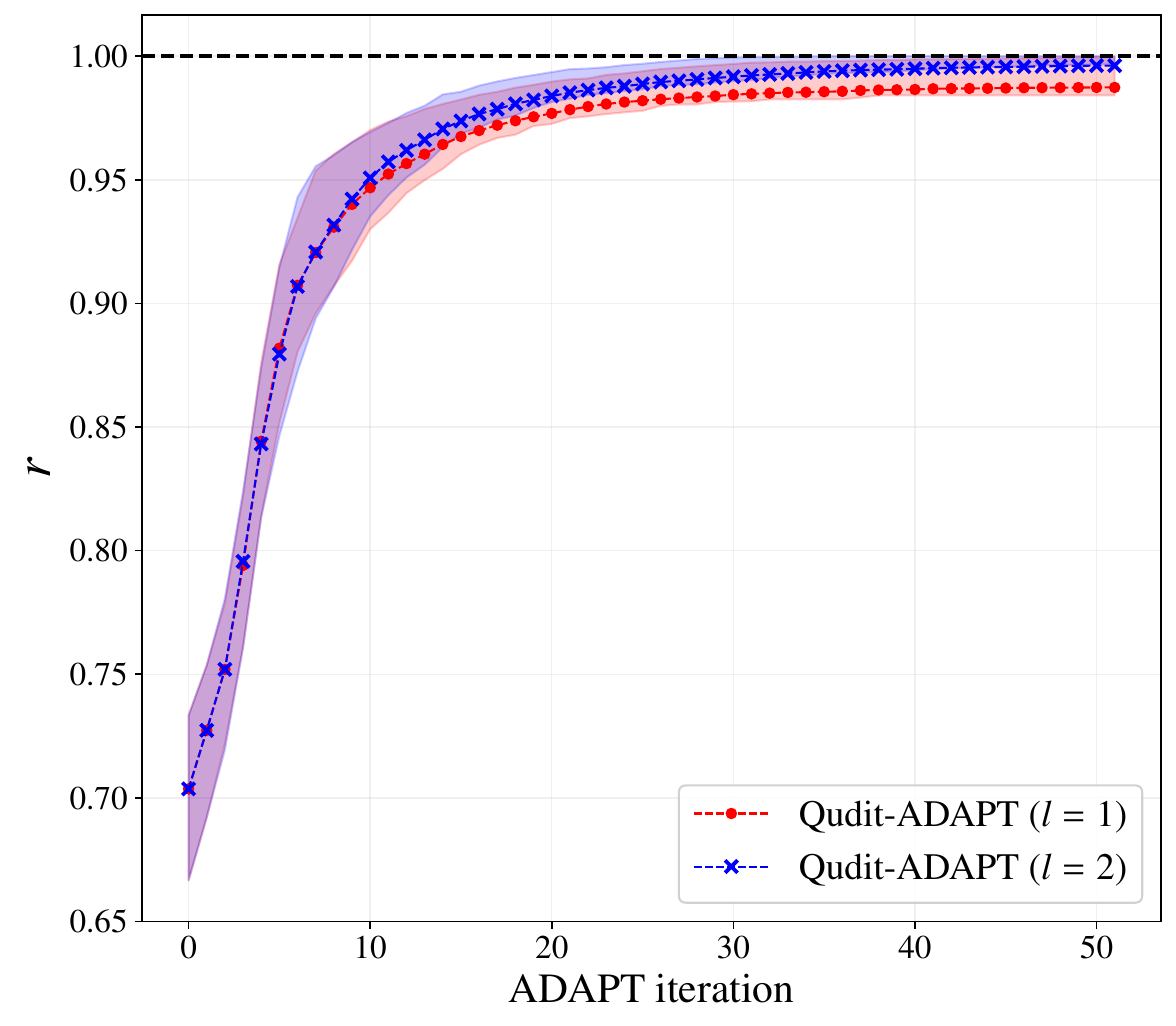}
            \put(2,88){b)}
        \end{overpic}
    \end{subfigure}

    \caption{a) Histogram of the approximation ratios obtained for all graph instances achieved by Qudit-ADAPT using the operator pools derived from the approximate AGP at orders $l=1$ and $l=2$, respectively. b) Median approximation ratio over all graph instances at each ADAPT iteration, up to the maximum number of iterations $k_{\max}=50$. The shaded regions represent the corresponding interquartile ranges. }
    \label{fig4}
\end{figure}
To assess the versatility of Qudit-ADAPT, we next consider 300 randomly generated graphs with six vertices and varying connectivity. Each instance is solved using operator pools derived from the approximate AGP at orders $l=1$ and $l=2$. In this case, we do not impose a convergence threshold.  Instead, we fix a maximum number of ADAPT iterations, corresponding to the maximum number of operators added to the ansatz. This allows the two operator pools to be compared under the same ansatz-growth budget, independently of their individual convergence behavior. 

Figure~\ref{fig4} a) shows the distribution of the approximation ratios obtained for the 300 graph instances considering maximum ADAPT iterations of $k_{\max}=50$. For $l=1$, the distribution is broader, with most values lying between approximately $0.9$ and $1$. In contrast, for $l=2$, the distribution is more concentrated near the optimal value, with the majority of the instances achieving approximation ratios above $0.99$, indicating that Qudit-ADAPT achieves consistently high performance in this problem independent of graph considered. 

An informative comparison can be made by examining the performance of the $l=1$ and $l=2$ operator pools throughout the ADAPT iterations. Figure~\ref{fig4} b) shows the median approximation ratio over the 300 random graph instances as a function of the ADAPT iteration, together with the corresponding interquartile ranges. This allows us to track the performance of both pools during the progressive construction of the ansatz.
During the first ADAPT iterations, the two operator pools exhibit very similar behavior, both in terms of the median approximation ratio and the interquartile range. This indicates that, at this stage, the operators selected from the $l=2$ pool, including those that are not present in the $l=1$ pool, do not yet provide a significant improvement in performance. A clear difference emerges after approximately 30 ADAPT iterations. Beyond this point, the $l=1$ results saturate, whereas the $l=2$ pool continues to improve. The absence of further improvement in the approximation ratio indicates that the gradient norm has become sufficiently small that, had a convergence threshold been imposed, the algorithm would have terminated at this point. The additional operators available at $l=2$ overcome this limitation and become increasingly relevant as the ansatz grows. Therefore, although both pools perform similarly during the initial stages of the ADAPT procedure, the additional operators contained in the $l=2$ pool become relevant at larger ansatz sizes and enable approximation ratios closer to the optimal value.

\subsection{Multiway Number Partitioning}

The Multiway Number Partitioning problem consists of dividing a set of $n$ real numbers into distinct subsets such that their sums are as balanced as possible. In this work, we focus on the case of three partitions, which can be naturally represented using qutrits.

Consider a set of real numbers defined as
\begin{equation}
    S = \{s_1, s_2, \ldots, s_n\},
\end{equation}
where $s_i \in \mathbb{R}$. In this work, we consider positive integers, $s_i \in \mathbb{N}$. 
To specify the partition to which each element belongs, we associate a ternary variable $v_i \in \{-1,0,1\}$ with each $s_i$. The three possible values correspond to the three partitions: $v_i = 1$ indicates that $s_i$ belongs to the first partition, $v_i = 0$ assigns it to the second partition, and $v_i = -1$ places it in the third partition. Consequently, the partitioning of the entire set can be encoded using $n$ qutrits, with each qutrit representing the assignment of a single element. Previous studies have explored this problem within the framework of qudit-based quantum computing \cite{TancaraNPJQuantumInf2025, BottarelliPRR2025}. Following these approaches, the associated cost Hamiltonian can be expressed as
\begin{equation}
\begin{aligned}
H_C = \frac{1}{4}\sum_{i,j=1}^{N}
\Bigg[
&6\left(L_z^{(i)}\right)^2\left(L_z^{(j)}\right)^2
+2L_z^{(i)}L_z^{(j)}
-4\left(L_z^{(i)}\right)^2 \\
&-4\left(L_z^{(j)}\right)^2
+\frac{8}{3}
\Bigg]s_i s_j.
\end{aligned}
\label{eq:hc_mwnp}
\end{equation}

Using $(L_z^{(i)})^4=(L_z^{(i)})^2$, each term of Eq.~\eqref{eq:hc_mwnp} with $i=j$ reduces to the constant $\frac{2}{3}s_i^2$, and each term with $i\neq j$ to $\frac{1}{2}s_is_j\,h_{ij}+\frac{2}{3}s_is_j$, where $h_{ij}=L_z^{(i)}L_z^{(j)}-2\big[(L_z^{(i)})^2+(L_z^{(j)})^2\big]+3(L_z^{(i)})^2(L_z^{(j)})^2$ is the edge term of the Max 3-Cut Hamiltonian in Eq.~\eqref{eq:maxcut_hc}. Adding all the terms gives
\begin{equation}
H_C = \sum_{i<j} s_i s_j\, h_{ij} + \frac{2}{3}\Big(\sum_{i=1}^{n} s_i\Big)^2.
\end{equation}
Partitioning into three subsets is therefore equivalent, up to a constant, to a weighted Max 3-Cut problem on the complete graph $K_n$ with edge weights $s_is_j$. 

\begin{figure}[H]
    \centering
    \includegraphics[width=0.99\linewidth]{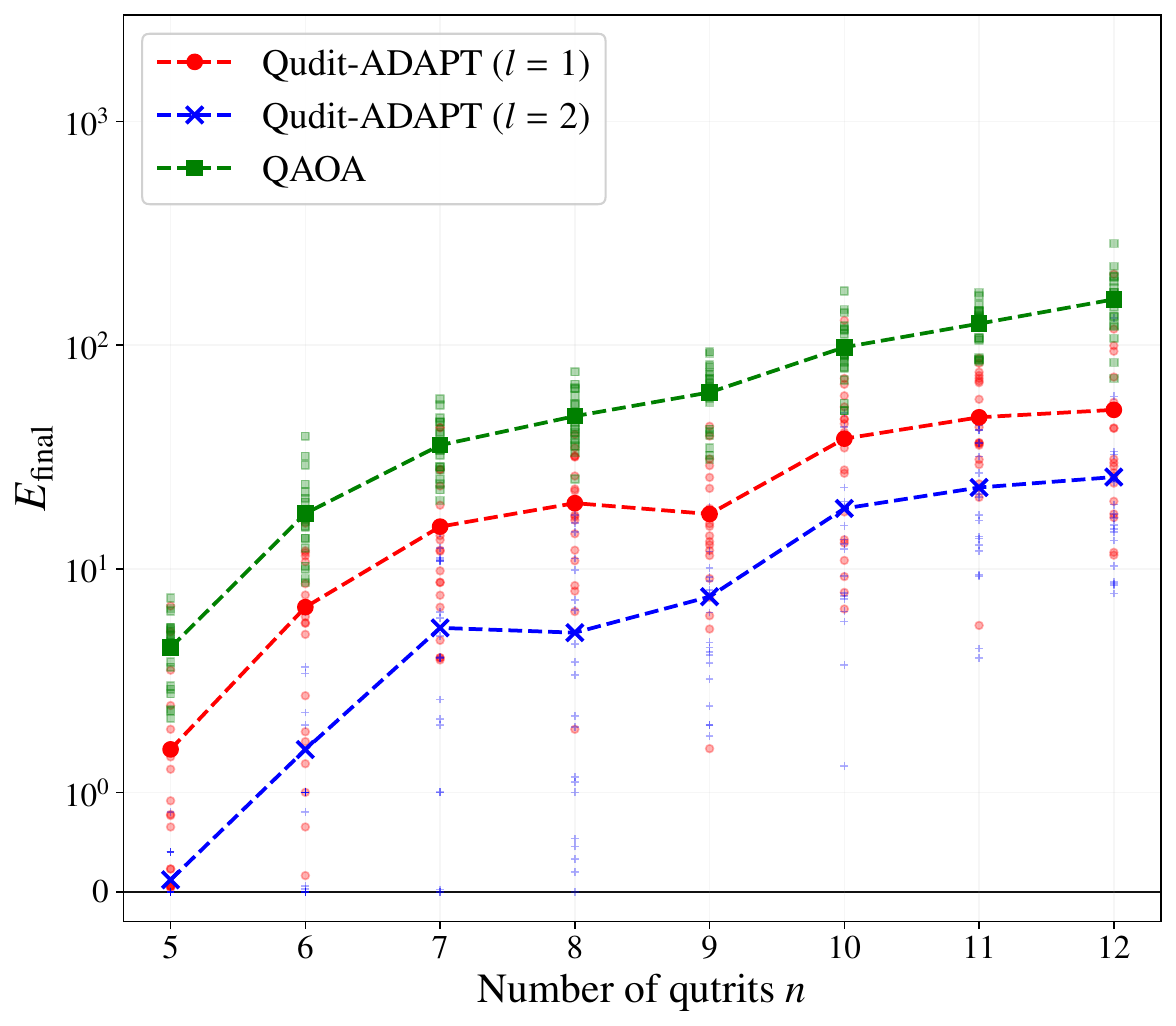}
    \caption{Final energy $E_{\mathrm{final}}$ as a function of the number of qutrits $n$ for the balanced multiway number partitioning problem, with a maximum of 80 variational parameters. Results are shown for Qudit-ADAPT with operator pools $l=1$ (red) and $l=2$ (blue), and for QAOA with $p=40$ layers initialized with the INTERP strategy (green). Small markers correspond to individual instances, 20 per system size, where the Qudit-ADAPT value is the average over the five orderings of the input numbers. Large markers connected by dashed lines are the averages over the 20 instances. Since $E_0=0$, the horizontal line marks the exact solution.}
    \label{fig_mwn2}
\end{figure}

To investigate the performance of Qudit-ADAPT as the system size increases, we consider balanced instances of the multiway number partitioning problem, with the number of qutrits ranging from $n=5$ to $n=12$. Specifically, we focus on instances that admit a perfect partition, where the sums of the three subsets are equal. For each system size, we generate 20 random instances consisting of distinct positive integers sampled from the interval $[1,3n]$. Representative instances are listed in Appendix~\ref{appendix_graphs}.

We compare Qudit-ADAPT with operator pools constructed up to orders $l=1$ and $l=2$ against QAOA, using the same budget of at most 80 variational parameters. Qudit-ADAPT terminates when the gradient norm falls below $\varepsilon=10^{-3}$ or when 80 operators have been added to the ansatz, and each instance is evaluated using five different orderings of the input numbers, one ascending ordering and four random permutations, for the reason discussed at the end of this section. For QAOA, we consider up to $p=40$ layers, corresponding to 80 variational parameters. Since the optimization of QAOA from random initial parameters becomes increasingly difficult as $p$ grows, we use the INTERP initialization strategy introduced in Ref.~\cite{ZhouPRX2020}. INTERP constructs the circuit layer by layer, starting from a single-layer QAOA optimized over multiple initializations. The optimal parameters at depth $p$ are then used to initialize the optimization at depth $p+1$, up to $p=40$, avoiding random initialization at each step. In our simulations, for $n\geq10$ the best of 10 QAOA optimizations from random parameters at $p=40$ barely lowers the energy of the initial state $\ket{\phi_g}$, whereas QAOA with INTERP reaches energies about 25 times lower.
\onecolumngrid

\begin{center}
    \includegraphics[width=0.95\textwidth]{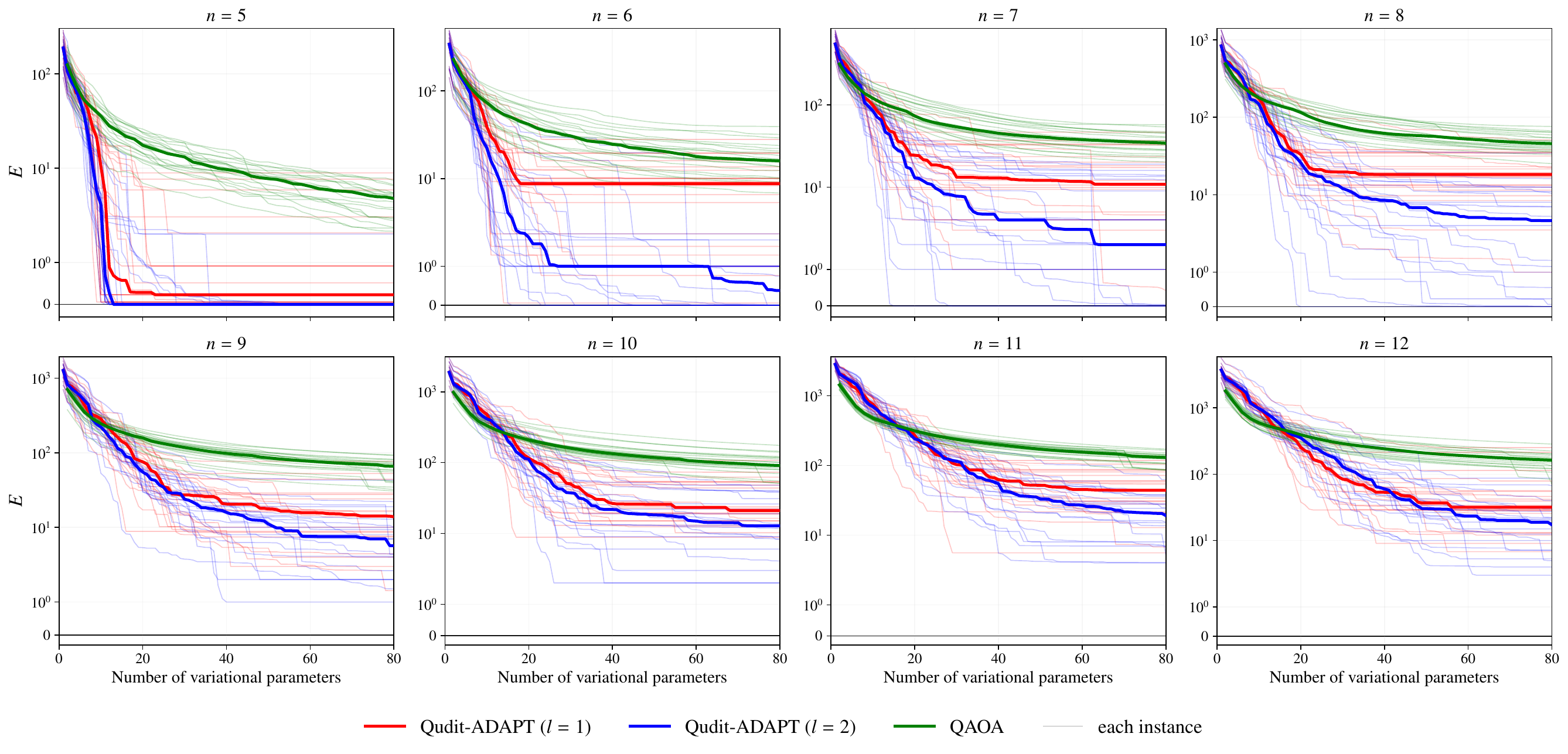}
    \captionof{figure}{Energy as a function of the number of variational parameters for the balanced multiway number partitioning problem with $n=5$ to $n=12$ qutrits. Results are shown for Qudit-ADAPT with operator pools $l=1$ (red) and $l=2$ (blue), and for QAOA initialized with the INTERP strategy (green), for which each layer adds two parameters. Thin lines correspond to individual instances, using the ascending ordering of the input numbers for Qudit-ADAPT, and thick lines indicate the median over the 20 instances. All methods are evaluated using up to 80 variational parameters.}
    \label{E_vs_n}
\end{center}

\twocolumngrid
Figure~\ref{fig_mwn2} presents the final energy $E_{\mathrm{final}}$ obtained by the different algorithms as a function of the number of qutrits. Since all considered instances admit a perfect partition, the ground-state energy is $E_0=0$, and therefore $E_{\mathrm{final}}$ directly measures the deviation from the optimal solution. Both versions of Qudit-ADAPT achieve lower energies than QAOA across all investigated system sizes. Qudit-ADAPT with $l=2$ reaches a lower final energy than QAOA in all 160 instances, and with $l=1$ in 155 of them, with average final energies between 5 and 37 times lower than QAOA for $l=2$ and between 2.3 and 3.5 times lower for $l=1$. The $l=2$ pool provides the lowest energies, outperforming $l=1$ in 142 of the 160 instances, which highlights the advantage of including higher-order operators in the ansatz. Although the final energies increase with the number of qutrits, they remain small compared with the energy of the initial state, $\bra{\phi_g}H_C\ket{\phi_g}$, reaching on average between $0.7\,\%$ and $2.2\,\%$ of it for $l=1$ and less than $1\,\%$ for $l=2$, compared with $1.9\,\%$ to $5.2\,\%$ for QAOA. We note, however, that preparing the exact solution, $E_{\mathrm{final}}=0$, becomes increasingly rare as $n$ grows. With the $l=2$ pool, $94\,\%$ of the runs reach $E_{\mathrm{final}}<10^{-3}$ for $n=5$, $11\,\%$ for $n=8$, and none for $n\geq10$ within the 80-parameter budget, which motivates the use of larger budgets or higher-order pools for larger instances.

To further examine the convergence behavior, Fig.~\ref{E_vs_n} shows the energy as a function of the number of variational parameters for each system size. Qudit-ADAPT exhibits a rapid initial decrease in energy, particularly for the $l=2$ pool, which generally reaches lower energies than the $l=1$ pool for the same number of parameters. In contrast, QAOA displays a slower reduction in energy and remains at higher values. Overall, these results demonstrate the effectiveness of adaptively selecting variational operators for solving balanced multiway number partitioning instances, compared with a fixed-structure QAOA ansatz under the same variational parameter budget.

Finally, the results of Qudit-ADAPT depend on the ordering of the input numbers, that is, on which number is assigned to each qutrit. Several operators in the pool can share the same maximum gradient magnitude. For example, at the first iteration the operators $L_y^{(i)}L_z^{(j)}$ and $L_z^{(i)}L_y^{(j)}$ acting on the same pair of qutrits have exactly the same gradient. In such cases the selection is resolved by the position of the operators in the pool, which follows the labeling of the qutrits, and differences at the level of floating-point rounding can decide between nearly degenerate gradients. Permuting the input numbers therefore leads to different sequences of variational operators and, consequently, different final results, which is why each instance is evaluated with five orderings. Developing a permutation-independent selection criterion to identify the most effective operator among those with equal maximum gradients remains an interesting direction for future work.

\subsection{Local minima and optimization landscapes}

A fundamental question in the design of Variational Quantum Algorithms (VQAs) is whether the ansatz is trainable, meaning that the variance of the cost function gradients decays at most polynomially with the system size, allowing an optimizer to converge efficiently~\cite{LaroccaNatRevPhys2025}. Recent analytical and numerical studies have shown that under certain conditions, VQAs suffer from \textit{barren plateaus} (BPs)~\cite{LaroccaNatRevPhys2025, HolmesPRXQuantum2022}, where the gradient variance vanishes exponentially and the optimization landscape becomes exponentially concentrated and thus flat. Even in the absence of BPs, the landscape can be highly rugged and densely populated with local minima, causing optimizers to stall despite sizable initial gradients, since training these algorithms is, in general, NP-hard~\cite{BittelPRL2021}.
\onecolumngrid
\begin{center}
    \vspace{0.8em}
    \begin{minipage}{0.47\textwidth}
        \centering
        \begin{overpic}[width=0.8\linewidth]{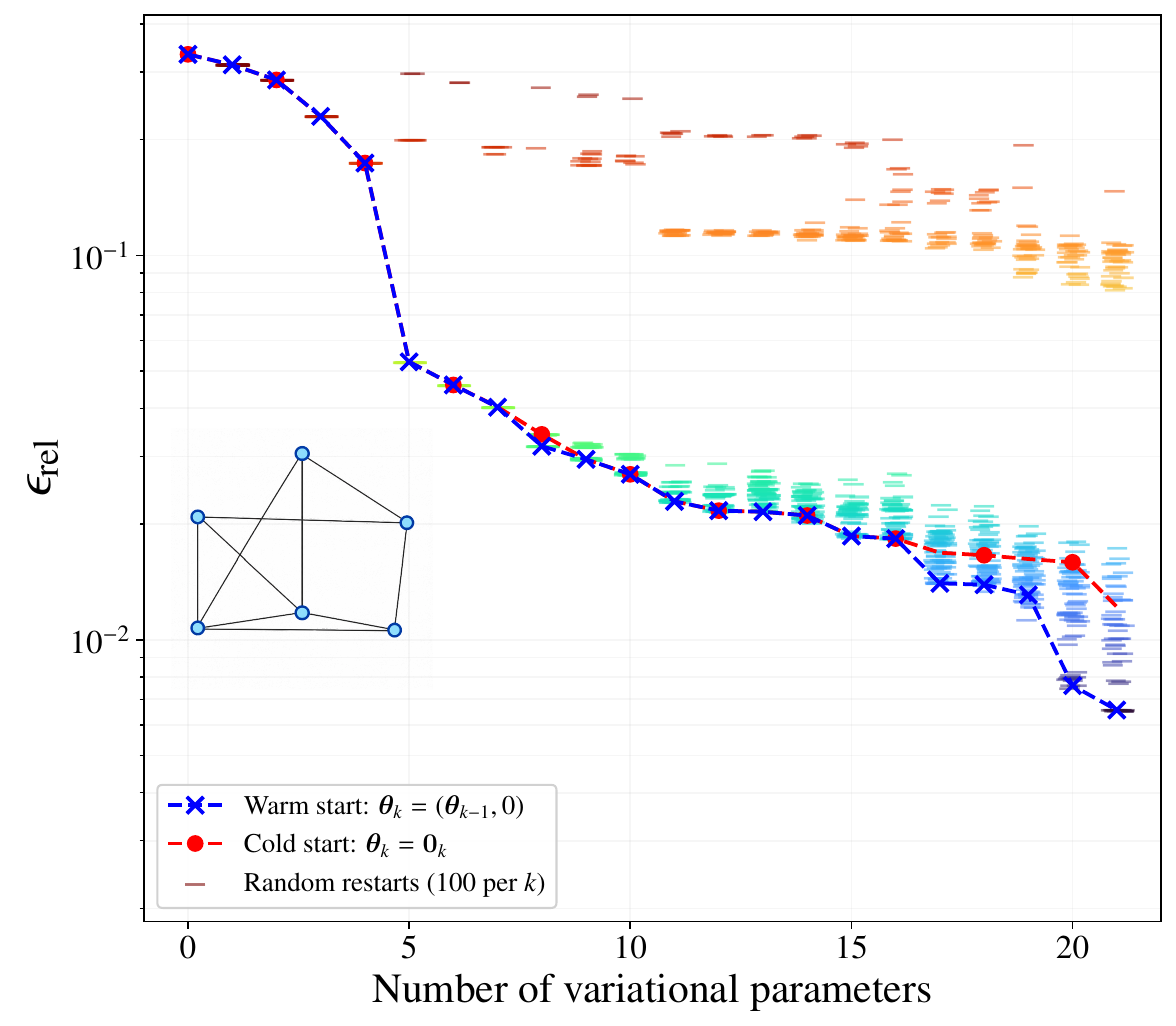}
            \put(2,88){a)}
        \end{overpic}

        \vspace{0.4em}

        \begin{overpic}[width=0.8\linewidth]{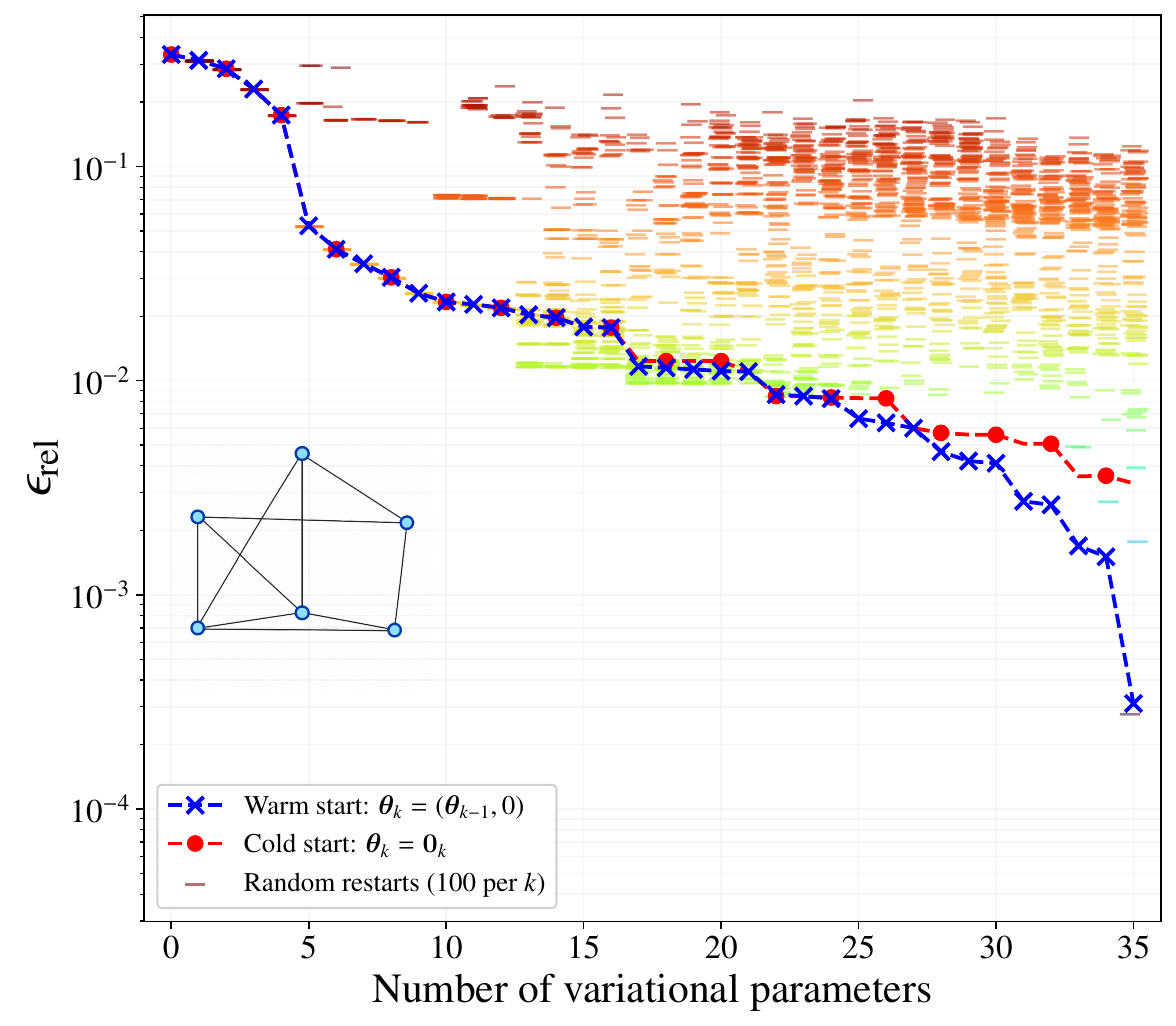}
            \put(2,88){c)}
        \end{overpic}
    \end{minipage}%
    \hfill
    \begin{minipage}{0.47\textwidth}
        \centering
        \begin{overpic}[width=0.8\linewidth]{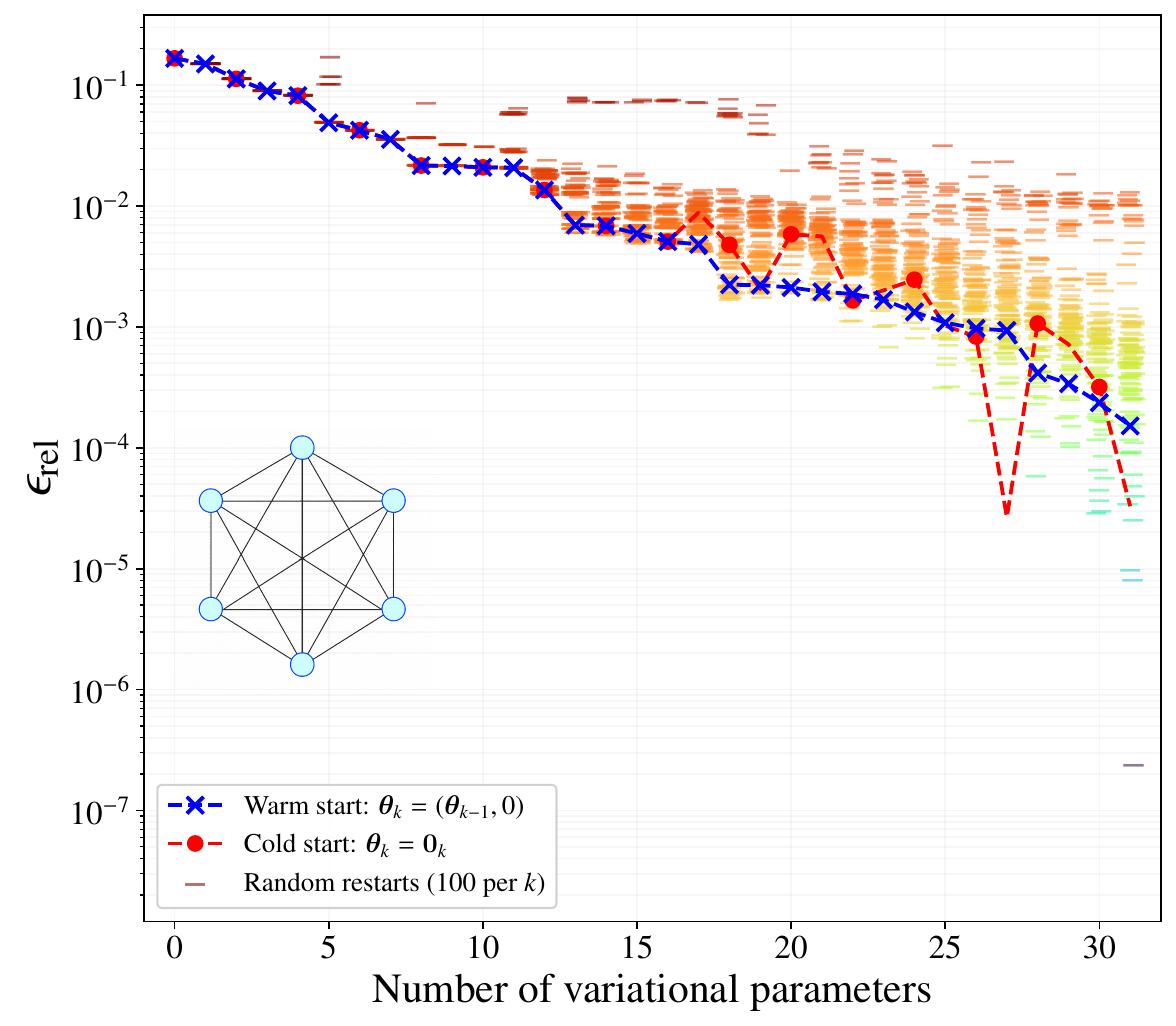}
            \put(2,88){b)}
        \end{overpic}

        \vspace{0.4em}

        \begin{overpic}[width=0.8\linewidth]{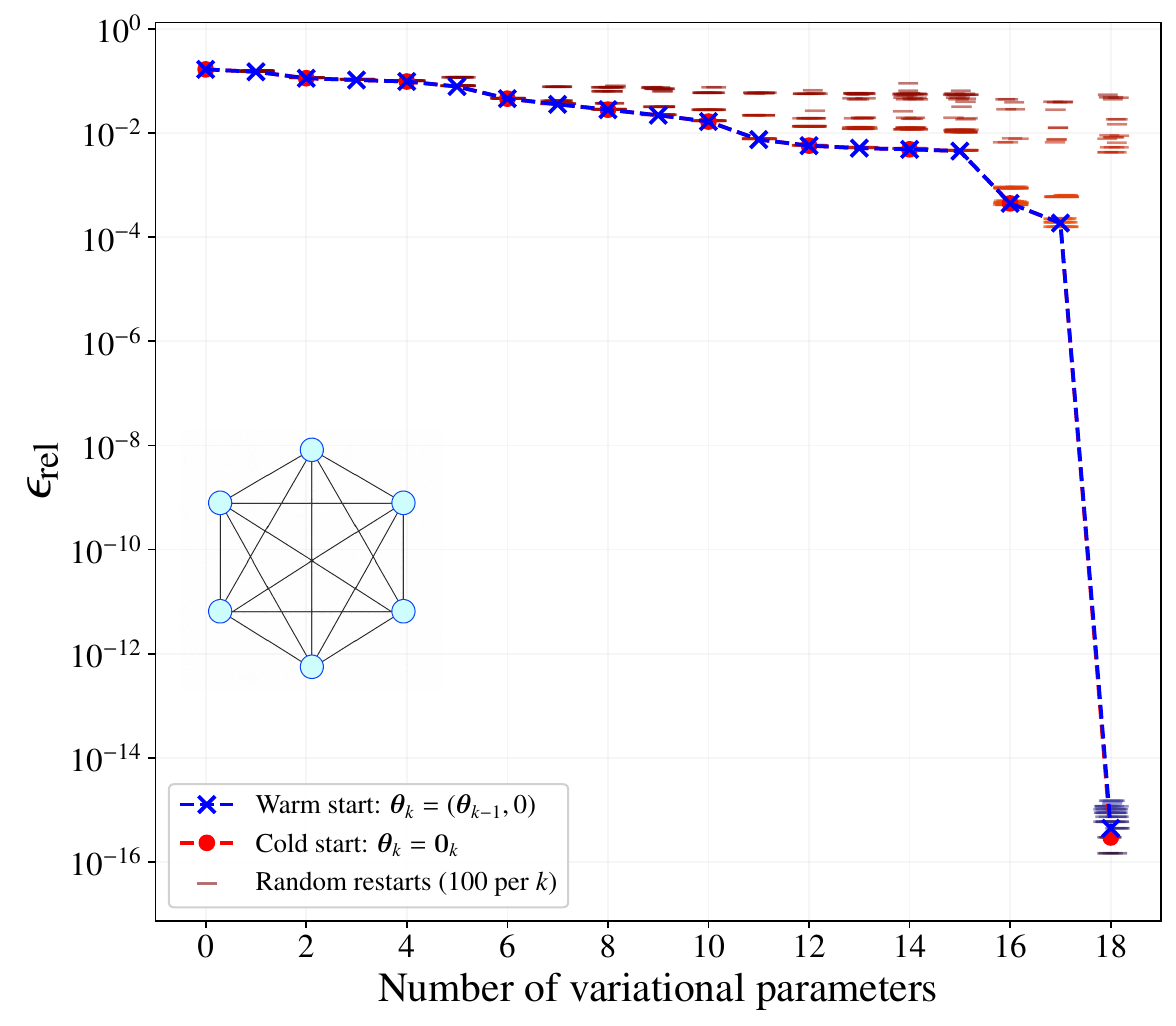}
            \put(2,88){d)}
        \end{overpic}
    \end{minipage}

    \captionof{figure}{Relative error as a function of the number of variational parameters, using operator pools obtained from the a,b) $l=1$ and c,d) $l=2$ truncations of the approximate AGP, for instance $G_1$ (a,c) and the complete graph $K_6$ (b,d). Each horizontal mark is the converged optimum of one of $100$ independent random parameter initializations of the same ansatz, colored by its relative error. The blue curve is the warm-start strategy used by Qudit-ADAPT and the red curve a cold restart at $\boldsymbol{\theta}_k=\boldsymbol{0}_k$. For $K_6$, the $l=1$ and $l=2$ operator pools contain $66$ and $2724$ operators, respectively; with the $l=2$ pool the algorithm converges to machine precision after $18$ parameters, where the warm-start and the cold restart coincide and $79$ of the $100$ random initializations reach the same minimum.}
    \label{fig_landscapes}
\end{center}
\twocolumngrid
The extension of VQAs to qudit architectures has deepened our understanding of how the Lie algebra of the operators involved in constructing the ansatz, and the estimation of the cost function and its gradients, influence optimization landscapes. Recent studies have shown that increasing the local Hilbert space dimension $d$ can amplify the barren plateau phenomenon ~\cite{FriedrichQMI2025}. In the regime of highly expressive qudit-based VQAs, where the ansatz approaches an approximate 2-design, the gradient variance scales as $\mathrm{Var}\{\partial_{\theta} C(\boldsymbol{\theta})\} \propto d^{-n}$~\cite{FriedrichQMI2025}, where $n$ is the number of qudits. Consequently, trainability is severely hindered in high-dimensional qudit systems unless specific mitigation strategies are employed. Qudit-ADAPT inherits the property that initial iterations correspond to shallow circuits, which therefore do not approximate 2-designs, avoiding barren plateaus at initialization and ensuring a resolvable gradient. Notably, ADAPT-VQE has shown favorable gradient behavior, while efficient classical simulability throughout its training remains an open question~\cite{Cerezo2025}.

As previously established for qubit systems~\cite{GrimsleynpjQuantumInf2023}, the ADAPT-VQE framework inherently mitigates these trainability obstructions through its iterative ansatz construction. To quantify this effect in Qudit-ADAPT, we isolate the specific contribution of the warm-start initialization. At every ADAPT iteration $k$, we re-optimize the same $k$-parameter ansatz from three distinct initial conditions:

\begin{enumerate}
    \item the \emph{warm-start} naturally provided by the algorithm, $\boldsymbol{\theta}_k = (\boldsymbol{\theta}_{k-1}^{\,*}, 0)$, where $\boldsymbol{\theta}_{k-1}^{\,*}$ are the optimized parameters from the previous iteration;
    \item a \emph{cold restart}, $\boldsymbol{\theta}_k = \boldsymbol{0}_k$, which resets the system to the reference state $\ket{\phi_g}$ at every step;
    \item a set of $100$ independent random initializations, where each parameter is sampled uniformly as $\theta_j \sim \mathcal{U}(-\pi,\pi)$.
\end{enumerate}

In all three cases the operator sequence is identically fixed to the one selected by the Qudit-ADAPT gradient criterion, so the optimization starting point is the only variable. All optimizations are performed with $\varepsilon=10^{-2}$ and a maximum of 35 iterations. For the random initializations we record only the converged energy of the optimizer. As a result, each data point in Fig.~\ref{fig_landscapes} represents a genuine local minimum where an optimization terminated, and each column provides a cross-sectional view of the local-minimum landscape characterizing an ansatz of that specific depth.

We first analyze the irregular graph instance $G_1$, shown in panels a) and c) of Fig.~\ref{fig_landscapes}. For $l=2$ the $100$ random initializations converge to $100$ distinct minima spanning more than two orders of magnitude in relative error, from $2.8\times10^{-4}$ to $1.3\times10^{-1}$. This is not a landscape in which an uninformed optimizer succeeds by chance, since the landscape is genuinely rough. The warm-start nevertheless descends monotonically throughout the run and reaches $\epsilon_{\mathrm{rel}} = 3.1\times10^{-4}$, against a median of $5.5\times10^{-2}$ over the random restarts, an improvement of more than two orders of magnitude with respect to the typical outcome of an uninformed optimization of the very same ansatz. Only $1$ of the $100$ restarts terminates below the warm-start value, and by a margin of about $10\,\%$. For $l=1$ the algorithm converges at $k=21$ with $\epsilon_{\mathrm{rel}} = 6.6\times10^{-3}$, against a median of $1.4\times10^{-2}$ and a worst case of $1.5\times10^{-1}$ among $88$ distinct minima. In this case, $6$ of the $100$ restarts reach the same minimum as the warm-start, within numerical precision, and none terminates below it.

Among the three initialization strategies considered, the comparison between the warm-start and the cold restart is particularly informative, since both employ identical operator sequences and pool choices and differ only in the parameter initialization. For $l=2$ the two curves overlap up to $k\simeq17$ and separate thereafter, with the cold restart saturating at approximately $\epsilon_{\mathrm{rel}}=3.3\times10^{-3}$ while the warm-start improves by a further order of magnitude. For $l=1$ the separation appears already at $k\simeq8$, and the cold restart ends a factor of two above the warm-start.

% ---------------------------------------------------------------------

As discussed in Ref.~\cite{GrimsleynpjQuantumInf2023}, ADAPT-VQE can ``burrow'' out of local minima by successively adding operators to the ansatz, thereby allowing further energy reduction even when the previous optimization converged to a local trap. Since the new parameter is initialized at zero, the energy can never increase from one iteration to the next, and the previous minimum ceases to be stationary as soon as an operator with a nonzero gradient is added. This mechanism does not prevent the algorithm from terminating at a suboptimal point when the gradients of the entire pool become small, which is what limits the $l=1$ results. Consistently, for $G_1$ with $l=1$ the variance of $\partial_\theta C$ evaluated at the random initial points varies only within a factor of $2.4$ over the whole run, between $1.6\times10^{-1}$ and $3.9\times10^{-1}$ from $k=1$ to $k=21$. What traps the random restarts is therefore the presence of local minima rather than a vanishing gradient. For $l=2$ the variance decreases from $3.9\times10^{-1}$ at $k=5$ to $4.1\times10^{-2}$ at $k=35$, consistent with the higher expressivity of the larger pool. Since these values correspond to a fixed system size, $n=6$, they do not by themselves probe the exponential decay with $n$ that defines a barren plateau.

The advantage depends on how expressive the pool is relative to the instance, and modifying the operator pool therefore changes the roughness of the parameter landscape itself. Panels b) and d) of Fig.~\ref{fig_landscapes} repeat the analysis for the complete graph $K_6$, whose pool contains $66$ operators at $l=1$ and $2724$ at $l=2$, compared with $46$ and $1138$ for $G_1$..

For $l=1$ the landscape remains rough, as the 100 random initializations converge to 99 distinct minima spanning almost five orders of magnitude, and the warm-start improves on their median by a factor of about $3.6$, $1.5\times10^{-4}$ against $5.4\times10^{-4}$. The comparison is nevertheless less clean than for the irregular graph, since $14$ of the $100$ random restarts terminate below the warm-start value and the cold restart ends a factor of $4.6$ lower. This is not in conflict with the burrowing argument, which guarantees monotonic improvement and the existence of a descent direction at every step, but not global optimality of the $k$-parameter landscape. Starting from $(\boldsymbol{\theta}_{k-1}^{\,*},0)$ keeps the optimization within the basin containing the previous optimum, so a deeper basin located elsewhere is unreachable without first increasing the energy, whereas a random initialization may land in it directly. Such an event, however, has low probability and costs a full optimization each time it is attempted. Recovering the best of $100$ restarts requires two orders of magnitude more optimizations than the single one performed by the warm-start, and without prior knowledge of $E_0$ there is no criterion to decide how many restarts are sufficient.

For $l=2$ the picture changes qualitatively. The algorithm reaches machine precision after $18$ parameters, the warm-start and the cold restart converge to the same solution, and so do $79$ of the $100$ random initializations, while the remaining $21$ terminate at relative errors between $4\times10^{-3}$ and $5\times10^{-2}$. Once the ansatz is sufficiently expressive, the global minimum becomes the dominant basin of the landscape, although a fraction of random initializations still terminates in suboptimal minima. Taken together, Fig.~\ref{fig_landscapes} indicates that the benefit of the warm-start is largest precisely in the regime of practical interest, where the pool is deliberately kept compact in order to limit circuit depth, and becomes less significant for highly connected and symmetric instances, where the larger operator pool contains sufficiently expressive terms to construct an accurate ansatz, indicating that the optimization problem for these instances is comparatively less challenging.

\section{Conclusions}
\label{sec:conclusions}
We propose an adaptive variational quantum algorithm for qudits that constructs the ansatz from an operator pool inspired by counterdiabatic driving, and apply it to several instances of the Max 3-Cut and Multiway Number Partitioning problems. To the best of our knowledge, this is the first work to employ the ADAPT-VQE strategy in qudit-based quantum computing. The Hamiltonians are formulated in terms of angular-momentum operators, whose closed commutation relations make them particularly convenient for the nested-commutator calculations required to construct the operator pool. Using numerical simulations, we implement Qudit-ADAPT with operator pools derived from approximate AGPs at orders $l=1$ and $l=2$ for different irregular and regular graph instances, and compare its performance with qudit QAOA. We observe that, with the $l=2$ pool, Qudit-ADAPT reduces the relative error by more than one order of magnitude compared with qudit QAOA for all four irregular benchmark instances. In general, Qudit-ADAPT also requires considerably fewer variational parameters and native gates to achieve a given level of accuracy than qudit QAOA.

To assess the generality of these results for Max 3-Cut, we additionally consider 300 graph instances. Approximation ratios above $0.9$ are obtained with the $l=1$ operator pool for the majority of instances, while $l=2$ achieves approximation ratios above $0.99$ for the majority of instances. An analysis of the approximation ratio throughout the ansatz construction shows that the $l=1$ and $l=2$ pools exhibit very similar performance during the first ADAPT iterations across the considered instances. At later iterations, however, the performance of the $l=1$ pool saturates, whereas the $l=2$ pool continues to improve and approaches unit approximation ratio. This behavior is mainly attributed to the greater expressivity provided by the additional operators contained in the $l=2$ pool.

We further investigated Qudit-ADAPT on balanced instances of the Multiway Number Partitioning problem with 5 to 12 qutrits. Under the same budget of 80 variational parameters, Qudit-ADAPT reaches lower final energies than qudit QAOA with INTERP initialization in all instances for $l=2$ and in 155 of the 160 instances for $l=1$, with the $l=2$ pool providing the best performance. Preparing the exact solution, however, becomes increasingly rare as the system size grows.

Finally, we analyze the behaviour under local minima in Qudit-ADAPT, motivated in part by the barren plateau problem, which has been shown to become more severe as the local Hilbert-space dimension increases. Representative irregular and regular graph instances are considered to assess the warm-start strategy against alternative parameter initializations, including a cold restart, in which all variational parameters are initialized to zero, and 100 random restarts. Similarly to the behavior reported in Ref.~\cite{GrimsleynpjQuantumInf2023}, we observe that the number of local minima tends to increase as more parameters are added to the ansatz. Nevertheless, the warm-start strategy consistently converges to energies below the median of the minima obtained from the random restarts. This behavior is consistent with the ``burrowing'' mechanism identified for ADAPT-VQE, whereby the addition of a new operator can deform the variational landscape and lower the local minimum reached by the previous ansatz, and we observe the same qualitative behavior in the qudit setting. We also find that the choice of operator pool changes the landscape itself. For the complete graph $K_6$, the $l=1$ ansatz presents a rough landscape with 99 distinct minima among 100 random initializations, whereas with the $l=2$ pool most initializations reach the exact solution. Together, these results suggest that the combination of warm-start initialization, iterative ansatz construction, and a sufficiently expressive operator pool provides robustness against local minima and may also mitigate optimization difficulties associated with barren plateaus.

In our approach, the expressivity of the operator pool can be systematically increased by considering higher-order approximations to the AGP. At the same time, several strategies originally developed for qubit-based ADAPT-VQE should be directly transferable to qudit implementations. Examples include incorporating hardware-aware penalties into the gradient-based operator-selection criterion~\cite{RamoaArxiv2026} and selecting groups of commuting operators to reduce the measurement cost associated with evaluating pool gradients~\cite{AnastasiouArxiv2023}. Such strategies may become particularly relevant as the size of the operator pool increases with the order of the approximate AGP.  For these reasons, we regard Qudit-ADAPT as a flexible framework for implementing variational quantum algorithms on qudit architectures, addressing two central challenges of fixed-ansatz approaches: ansatz design and robustness against unfavorable optimization landscapes, including local minima. 

\section{Data and Code Availability}

The data generated with the code that support the findings of this study are available publicly in this github repository: \href{https://github.com/JoaquinQu/Qudit-ADAPT}{https://github.com/JoaquinQu/Qudit-ADAPT}

\section{ACKNOWLEDGMENTS}
DT acknowledges grant Posdoctorado UC PD2024-609. DG and JM acknowledge grant FONDECyT Regular nr 1230586, Chile. HD acknowledges ANID BECAS/MAGÍSTER NACIONAL 22251911. HD and DG acknowledge grant Anillo ATE250066, Chile. ChatGPT and Claude were used to assist with language editing, improving the clarity of the manuscript, and refining the computational code. All scientific content, numerical results, calculations, and conclusions were reviewed and verified by the authors, who take full responsibility for the final manuscript.

\section{AUTHOR CONTRIBUTIONS}
JM and HD performed all numerical and analytical calculations and prepared the figures. DG supervised the work and reviewed the manuscript. DT conceived the original idea, verified the calculations, and wrote the manuscript.

\appendix
\section{Explicit form and Hermiticity of the operator pool}
\label{appendix1}

For the Max 3-Cut cost Hamiltonian, the first-order generator of the AGP expansion is
\[
O_1=[H_{\mathrm{ad}},O_0]=[H_M,H_C],
\]
since the $\lambda$-dependence cancels out between $[H_M,H_C]$ and $[H_C,H_M]$ in $H_{\mathrm{ad}}$, making $O_1$ independent of $\lambda$. Writing $H_M=-\sum_mX_m$ with $X_m=(L_z^{(m)})^2+\sqrt{2}L_x^{(m)}$ and $H_C=\sum_{{i,j}\in E}h_{ij}$, each $X_m$ acts only on qudit $m$ while each $h_{ij}$ acts only on qudits $i,j$; if $m\notin{i,j}$, $X_m$ and $h_{ij}$ act on different tensor factors and commute trivially, so only $X_i,X_j$ have a nonvanishing commutator with $h_{ij}$. Since $(L_z^{(i)})^2$ commutes with every term of $h_{ij}$, only the $L_x^{(i)}$ piece of $X_i$ contributes; using
\[
[L_x^{(i)},L_z^{(i)}]=-iL_y^{(i)},\qquad [L_x^{(i)},(L_z^{(i)})^2]=-i\{L_y^{(i)},L_z^{(i)}\},
\]
together with the analogous relations for site $j$,
\begin{align}
O_1={}&i\sqrt{2}\sum_{\{i,j\}\in E}\Big[L_y^{(i)}L_z^{(j)}+L_z^{(i)}L_y^{(j)}\notag\\
&-2\big(\{L_y^{(i)},L_z^{(i)}\}+\{L_y^{(j)},L_z^{(j)}\}\big)\notag\\
&+3\big(\{L_y^{(i)},L_z^{(i)}\}(L_z^{(j)})^2+(L_z^{(i)})^2\{L_y^{(j)},L_z^{(j)}\}\big)\Big].
\end{align}

In practice, each monomial appearing in $O_1$ is extracted individually as an operator string $D_j$ to build the operator pool [Eq.~\eqref{eq:agp_generator_expansion}]. Hermiticity must therefore be examined at the level of individual monomials, rather than for $O_1$ as a whole. Every operator string appearing here falls into one of two classes. Strings built from operators on different qudits, such as $L_y^{(i)}L_z^{(j)}$, are Hermitian on their own, since operators on different tensor factors commute and therefore equal their own adjoint. Strings built from two operators on the same qudit are not Hermitian individually, but appear together with their exact Hermitian conjugate at equal weight,
\[
(L_y^{(i)}L_z^{(i)})^\dagger=L_z^{(i)}L_y^{(i)},
\]
the other term inside $\{L_y^{(i)},L_z^{(i)}\}$. Consequently, Hermitizing the individual operator strings $D_j$ obtained from $O_1$ through $V_j=(D_j+D_j^\dagger)/2$ does not introduce any generator absent from the natural algebraic structure of the AGP: for same-qudit strings, $D_j^\dagger$ is already present in $O_1$ with the same coefficient, so $V_j$ only recombines terms the nested-commutator expansion already produces; for different-qudit strings, $V_j=D_j$ trivially.

This cancellation is not special to first order. Writing $O_k=\sum_jc_j^{(k)}D_j$ for the raw expansion at any order $k$, two general facts hold. First,
\[
O_k^\dagger=(-1)^kO_k,
\]
since $O_0=H_C-H_M$ is Hermitian and each commutator with the Hermitian $H_{\mathrm{ad}}$ turns a Hermitian operator anti-Hermitian and vice versa. Second, every nonzero coefficient satisfies
\[
c_j^{(k)}=i^k\rho_j,\qquad \rho_j\in\mathbb{R},
\]
since the only source of $i$ in the recursion is the single-qudit algebra $[L_a^{(i)},L_b^{(i)}]=i\epsilon_{abc}L_c^{(i)}$, invoked once per resolved commutator, while every other factor ($\lambda$, $1-\lambda$, $\sqrt2$, integer multiplicities) is real. Together these give $\hat{O}_k:=i^kO_k$ Hermitian with real coefficients $c_j:=i^k c_j^{(k)}=\rho_j$. Splitting each $D_j=V_j+W_j$ into Hermitian and anti-Hermitian parts,
\[
\hat{O}_k=\sum_jc_jV_j+\sum_jc_jW_j,
\]
and since this decomposition is unique for any operator while $\hat{O}_k$ is itself Hermitian, the anti-Hermitian piece must vanish, leaving
\[
\hat{O}_k=\sum_jc_jV_j
\]
exactly. Hermitizing therefore never discards nor introduces anything beyond $O_{2l-1}$'s own expansion, at any pool order $l$.

\section{Gate cost of the algorithms}
\label{app:gate_cost}

The feasibility of running a variational algorithm on current qudit hardware depends on the number of native operations of its circuits, since each operation adds to the accumulated error. To assess how close Qudit-ADAPT and QAOA are to an implementation on qutrit processors, we estimate the native gate cost of their circuits on the trapped-ion qudit processor of Ref.~\cite{RingbauerNatPhys2022}, for Max 3-Cut on $G_2$ and $G_5$ and for the first Multiway Number Partitioning (MWNP) problem instance with $n=6$, $S=\{1,6,8,10,12,17\}$, using the $l=1$ pool for Qudit-ADAPT. In this processor, single-qudit operations are laser pulses that drive a pair of levels $(a,b)$ of one ion, phases on a single level are tracked in software as virtual rotations at no cost, and two-qudit interactions are generated by M\o{}lmer-S\o{}rensen (MS) gates,
\begin{align}
\label{eq:pulse_ms}
R^{(a,b)}(\theta,\varphi) & =e^{-i\frac{\theta}{2}\sigma_\varphi^{(a,b)}},\\
\mathrm{MS}^{(a,b)}(\theta, \varphi) & =e^{-i\frac{\theta}{4}\left(\sigma_\varphi^{(a,b)}\otimes I+I\otimes\sigma_\varphi^{(a,b)}\right)^2},
\end{align}
where $\sigma_\varphi^{(a,b)}=\cos\varphi\,\sigma_x^{(a,b)}+\sin\varphi\,\sigma_y^{(a,b)}$, with $\sigma_x^{(a,b)}=\ketbra{a}{b}+\ketbra{b}{a}$, $\sigma_y^{(a,b)}=-i\ketbra{a}{b}+i\ketbra{b}{a}$ and $\sigma_z^{(a,b)}=\ketbra{a}{a}-\ketbra{b}{b}$. The three levels of each ion are labeled by the states $\ket{1}$, $\ket{0}$ and $\ket{{-1}}$ of Eq.~\eqref{eq:basis}, so that $a,b\in\{1,0,-1\}$. We compile the circuits with the open-source compiler MQT Qudits~\cite{MatoArxiv2024}, which decomposes one- and two-qudit unitaries into local pulses, virtual rotations and the controlled-exchange gate introduced in Ref.~\cite{RingbauerNatPhys2022}. This gate exchanges the levels $t_1$ and $t_2$ of the target qudit if and only if the control qudit is in level $c$. MQT Qudits implements it as a $\pi$ pulse conditioned on the control,
\begin{equation}
\label{eq:cex}
\mathrm{CEx}^{(c;t_1,t_2)}(\varphi)=\sum_{k\neq c}\ketbra{k}{k}\otimes I+\ketbra{c}{c}\otimes R^{(t_1,t_2)}(\pi,\varphi),
\end{equation}
which differs from the gate of Ref.~\cite{RingbauerNatPhys2022} by the phases $-ie^{\pm i\varphi}$ acquired by the two exchanged states.

For Qudit-ADAPT, each operator of the final ansatz, $e^{-i\theta_kV_{j_k}}$, is evaluated at its optimized angle and decomposed with MQT Qudits. For QAOA, the same procedure is applied to the cost and mixer unitaries of each layer, $e^{-i\gamma_jH_C}$ and $e^{-i\beta_jH_M^{\mathrm{QAOA}}}$, evaluated at their optimized parameters. Since the terms of each Hamiltonian commute, these unitaries factorize exactly into two-qutrit gates, one per edge, and single-qutrit gates, one per qutrit, which are decomposed one by one. The gates of every decomposed operator are then counted, and the counts are added. Evaluating each operator at its optimized angle is required because MQT Qudits decomposes numerical unitaries rather than parametrized gates. The counts do not depend on the angles except at special values, at which some rotations vanish. Single-qutrit unitaries are decomposed by a QR decomposition into two-level rotations~\cite{BullockPRL2005}, and two-qutrit unitaries by the QR decomposition of the $9\times9$ matrix, in which every two-level rotation becomes a rotation controlled by the other qutrit, built from CEx gates and local pulses~\cite{MatoASPDAC2023}. For QAOA we use $p=\lfloor P/2\rfloor$ complete layers, with $P$ the number of parameters of Qudit-ADAPT, and the angles of optimized QAOA runs with $p$ layers.

The compilation passes of the current release of MQT Qudits (0.5.2) do not reproduce the target unitary of non-diagonal two-qutrit gates. We therefore use the version of an open pull request of the MQT Qudits repository, not yet merged, which corrects this~\cite{MQTQuditsPR69}, with the two-qutrit trapped-ion test backend of the package. Numerical comparison of the compiled circuits with their corresponding exact target unitaries yields a maximum entrywise deviation of $4\times10^{-5}$, after accounting for global phases.

\begin{table}[H]
\centering
\resizebox{\columnwidth}{!}{%
\begin{tabular}{llccc}
\hline\hline
Instance & Method & Parameters & Local pulses $R^{(a,b)}$ & $\mathrm{CEx}$ \\
\hline
$G_2$ & Qudit-ADAPT ($l=1$) & 11 & 598 & 104 \\
      & QAOA ($p=5$) & 10 & 7074 & 1584 \\
\hline
$G_5$ & Qudit-ADAPT ($l=1$) & 12 & 638 & 112 \\
      & QAOA ($p=6$) & 12 & 7740 & 1728 \\
\hline
MWNP ($n=6$) & Qudit-ADAPT ($l=1$) & 13 & 619 & 108 \\
      & QAOA ($p=6$) & 12 & 14292 & 3240 \\
\hline\hline
\end{tabular}%
}
\caption{Compiled gates of Qudit-ADAPT ($l=1$) and QAOA, obtained by decomposing each operator of the final circuits with MQT Qudits and adding the counts. Virtual rotations are not counted, and the preparation of the initial state is not included.}
\label{tab:gate_cost}
\end{table}

Table~\ref{tab:gate_cost} shows that QAOA, with at most as many parameters, requires between $15$ and $30$ times more CEx gates and between $12$ and $23$ times more local pulses than Qudit-ADAPT. This reduction arises because each exponential of Qudit-ADAPT contains a single operator, whereas each exponential of QAOA contains a Hamiltonian with one term per edge or per qutrit, all of which must be implemented. The same behavior was observed when comparing the gate cost of CD-ADAPT with that of a digitized counterdiabatic protocol in molecular simulations~\cite{TancaraJCTC2026}.

We emphasize that these counts are not optimized, since the decomposition algorithms are designed for generic $\mathrm{SU}(d)$ unitaries and ignore the structure of the target operators. They are therefore upper bounds for both algorithms. For example, MQT Qudits compiles $e^{-i\theta L_z\otimes L_z}$ with $24$ CEx gates, whereas a single MS gate and four local pulses suffice. Since $L_z=\sigma_z^{(1,-1)}$ and $L_z^2=\ketbra{1}{1}+\ketbra{{-1}}{{-1}}$,
\begin{equation}
\label{eq:lzlz}
\begin{split}
e^{-i\theta L_z\otimes L_z}= &\big[R^{(1,-1)}(\tfrac{\pi}{2},\tfrac{\pi}{2})\,e^{i\frac{\theta}{2}L_z^2}\big]^{\otimes2}\,\mathrm{MS}^{(1,-1)}(2\theta,0)\\
\quad \times &\big[R^{(1,-1)}(-\tfrac{\pi}{2},\tfrac{\pi}{2})\big]^{\otimes2},
\end{split}
\end{equation}
where the diagonal phase $e^{i\frac{\theta}{2}L_z^2}$ is implemented as a virtual rotation. Structure-aware compilation would lower the cost of both algorithms, and possibly their ratio, and is left as a future work. Finally, MQT Qudits compiles operations on at most two qudits, and the compilation of multi-qudit operations is planned for future releases~\cite{MatoThesis2025}. Since the $l=2$ pool contains generators acting on three and four qutrits, its circuits could not be compiled with this tool.

\section{Problem instances}
\label{appendix_graphs}

\subsection{Max 3-Cut graphs}

The edge sets of the example graphs considered in this manuscript are

\onecolumngrid
\begin{align}
G_1 &= \{ (1,2), (1,3), (1,4), (1,5), (2,4), (2,6), (3,4), (3,6), (4,5), (5,6) \}, \\
G_2 &= \{ (1,2), (1,3), (1,4), (1,5), (2,3), (2,4), (2,5), (3,4), (3,5), (4,5), (5,6) \}, \\
G_3 &= \{ (1,2), (1,3), (1,5), (2,3), (2,4), (2,6), (3,4), (3,5), (3,6), (4,5), (5,6) \}, \\
G_4 &= \{ (1,2), (1,3), (1,4), (1,6), (2,3), (2,4), (2,5), (3,4), (3,5), (4,6), (5,6) \}, \\
G_5 &= \{ (1,2), (1,3), (1,4), (1,5), (2,3), (2,4), (2,6), (3,5), (3,6), (4,5), (4,6), (5,6) \}, \\
\begin{split}
K_6 = \{ & (1,2), (1,3), (1,4), (1,5), (1,6), (2,3), (2,4), (2,5), \\
        & (2,6), (3,4), (3,5), (3,6), (4,5), (4,6), (5,6) \}
\end{split}
\end{align}
\twocolumngrid

Instances $G_1$ to $G_4$ are the irregular benchmark graphs, and $G_5$ is $4$-regular while $K_6$ is the complete, $5$-regular graph on six vertices; all are shown in Fig.~\ref{fig_instances}.

\subsection{Multiway number partitioning instances}

For each system size $n=5,\ldots,12$, we generated 20 instances by sampling $n$ distinct integers from the interval $[1,3n]$ and retaining only those that admit a unique perfect partition into three subsets, up to permutations of the subsets. Table~\ref{tab:mwnp_instances} lists two representative instances for each system size together with their optimal partition. The complete set of 160 instances is available in the public repository associated with this work.

\onecolumngrid
\begin{center}
    \vspace{0.8em}
\begin{tabular}{cllc}
\hline\hline
$n$ & Set $S$ & Optimal partition & Subset sum \\
\hline
5 & $\{2,3,7,8,10\}$ & $\{10\}\;|\;\{2,8\}\;|\;\{3,7\}$ & 10 \\
5 & $\{1,4,6,9,10\}$ & $\{10\}\;|\;\{1,9\}\;|\;\{4,6\}$ & 10 \\[2pt]
6 & $\{1,6,8,10,12,17\}$ & $\{1,17\}\;|\;\{6,12\}\;|\;\{8,10\}$ & 18 \\
6 & $\{4,6,7,8,11,18\}$ & $\{18\}\;|\;\{7,11\}\;|\;\{4,6,8\}$ & 18 \\[2pt]
7 & $\{1,4,6,10,13,16,19\}$ & $\{4,19\}\;|\;\{10,13\}\;|\;\{1,6,16\}$ & 23 \\
7 & $\{3,5,6,9,11,17,18\}$ & $\{5,18\}\;|\;\{6,17\}\;|\;\{3,9,11\}$ & 23 \\[2pt]
8 & $\{8,9,10,11,12,14,16,19\}$ & $\{14,19\}\;|\;\{8,9,16\}\;|\;\{10,11,12\}$ & 33 \\
8 & $\{5,6,8,9,14,16,17,24\}$ & $\{9,24\}\;|\;\{16,17\}\;|\;\{5,6,8,14\}$ & 33 \\[2pt]
9 & $\{1,2,5,6,7,8,20,21,26\}$ & $\{6,26\}\;|\;\{5,7,20\}\;|\;\{1,2,8,21\}$ & 32 \\
9 & $\{2,3,4,6,8,9,12,23,26\}$ & $\{8,23\}\;|\;\{2,3,26\}\;|\;\{4,6,9,12\}$ & 31 \\[2pt]
10 & $\{2,3,12,15,16,17,18,22,27,30\}$ & $\{2,22,30\}\;|\;\{12,15,27\}\;|\;\{3,16,17,18\}$ & 54 \\
10 & $\{2,9,18,22,23,24,25,26,28,30\}$ & $\{18,25,26\}\;|\;\{22,23,24\}\;|\;\{2,9,28,30\}$ & 69 \\[2pt]
11 & $\{10,11,17,18,19,22,24,26,29,31,33\}$ & $\{18,29,33\}\;|\;\{10,17,22,31\}\;|\;\{11,19,24,26\}$ & 80 \\
11 & $\{8,14,15,16,18,22,25,26,29,31,33\}$ & $\{15,31,33\}\;|\;\{8,16,26,29\}\;|\;\{14,18,22,25\}$ & 79 \\[2pt]
12 & $\{8,9,11,12,17,18,19,24,26,33,34,35\}$ & $\{8,17,24,33\}\;|\;\{9,12,26,35\}\;|\;\{11,18,19,34\}$ & 82 \\
12 & $\{2,4,5,8,16,20,21,29,32,33,34,36\}$ & $\{2,16,29,33\}\;|\;\{4,8,32,36\}\;|\;\{5,20,21,34\}$ & 80 \\
\hline\hline
\end{tabular}
\captionof{table}{Two representative instances of the balanced multiway number partitioning problem for each number of qutrits $n$. Each set $S$ consists of $n$ distinct integers in $[1,3n]$ and admits a unique perfect partition into three subsets with equal sums, shown in the third column.}
\label{tab:mwnp_instances}
\end{center}
\twocolumngrid

\end{document}